# Methods for generating and evaluating synthetic longitudinal patient data: a systematic review

Katariina Perkonoja *

Department of Mathematics and Statistics, University of Turku, Finland

Department of Computing, University of Turku, Finland

Kari Auranen

Department of Mathematics and Statistics, University of Turku, Finland

Department of Clinical Medicine, University of Turku, Finland

Joni Virta

Department of Mathematics and Statistics, University of Turku, Finland

The rapid growth in data availability has facilitated research and development, yet not all industries have benefited equally due to legal and privacy constraints. The healthcare sector faces significant challenges in utilizing patient data because of concerns about data security and confidentiality. To address this, various privacy-preserving methods, including synthetic data generation, have been proposed. Synthetic data replicate existing data as closely as possible, acting as a proxy for sensitive information. While patient data are often longitudinal, this aspect remains underrepresented in existing reviews of synthetic data generation in healthcare. This paper maps and describes methods for generating and evaluating synthetic longitudinal patient data in real-life settings through a systematic literature review, conducted following the PRISMA guidelines and incorporating data from five databases up to May 2024. Thirty-nine methods were identified, with four addressing all challenges of longitudinal data generation, though none included privacy-preserving mechanisms. Resemblance was evaluated in most studies, utility in the majority, and privacy in just over half. Only a small fraction of studies assessed all three aspects. Our findings highlight the need for further research in this area.

## 1 INTRODUCTION

The recent surge in data volumes has greatly facilitated research, development, and innovation (RDI) activities. Yet, some sectors, particularly medicine, still face challenges in harnessing existing data sources due to stringent data protection regulations, such as the General Data Protection Regulation [84] or the Health Insurance Portability and Accountability Act [83]. Compliance with these policies leads to prolonged data processing times and, in certain cases, restricted access. For instance, while the national regulation in Finland permits using identifiable individual-level data for research, their use in development and innovation activities remain prohibited [17].

If patient data are deemed anonymous, they fall outside the rules of personal data protection, streamlining data access and sharing. Synthetic data generation (SDG), originally proposed by Rubin [56] in 1993, offers a promising approach to achieve such anonymity. The goal of SDG is to produce artificial data that resemble real-world observations, referred to as original or input data, while maintaining adequate utility and resemblance with the original. Here, resemblance refers to similarity between the synthetic and original data distributions, while utility pertains to the extent the analyses and predictions based on the synthesized data align with those from the original data. However, concerns regarding the sufficiency of mere random data generation for privacy preservation have prompted exploration of more effective privacy-preserving techniques in SDG [55, 65]. Beyond privacy, synthetic data (SD) help establish analytical environments and model testing in preparation for real data, while also enabling data augmentation for underrepresented observations (class imbalance). Consequently, SDG has gained prominence in enhancing data protection and facilitating RDI activities as well as educational purposes.

---

* Corresponding author. K.P: Conceptualization, Methodology, Formal analysis, Investigation, Resources, Data curation, Writing – original draft, Visualization, Project administration, Funding acquisition. K.A: Conceptualization, Methodology, Supervision, Writing – review & editing. J.V: Conceptualization, Methodology, Investigation, Writing – review & editing, Supervision, Funding acquisition. The authors declare no potential conflicts of interest with respect to the research, authorship, and/or publication of this article.

Medical data encompass various forms, with longitudinal data (LD) being particularly valuable for their ability to provide detailed insights into patient trajectories over time. LD consist of at least one variable measured for each subject at two or more time points [13]. In addition to these time-varying repeated measurements, LD often include static variables, such as patient demographics. The integration of static and time-varying variables in LD enables a more comprehensive analysis of patient well-being, offering improved potential for personalized and preventive medicine. These insights extend beyond what can be derived from cross-sectional data that capture information only at a single time point [13].

When repeated measurements are collected uniformly across subjects, LD are considered *balanced*. Conversely, if the timing or frequency of measurements varies between subjects, the LD are called *unbalanced*. The question of unbalancedness is distinct from that of missing data; in an unbalanced dataset, the reasons for irregular observations are known, whereas with missing data, also the cause and nature of the missing values must be examined [14]. Notably, missing data can occur in both balanced and unbalanced LD structures. Longitudinal data can be collected retrospectively, such as from electronic health records (EHRs), or prospectively, as in clinical trials and cohort studies. Unbalanced data often arise in secondary use due to individualized patient treatment, whereas prospective studies typically follow planned measurement schedules, resulting in more balanced data. Nonetheless, unbalance can occur in prospective studies, for instance, when cost constraints reduce measurement frequency for certain participants. Figure 1 depicts the key characteristics of longitudinal data.

**Balanced longitudinal data**

| Patient ID | Sex | Group | Intervention | Age | Height | Weight | CCI | Albumin | CRP | QoL | ··· | Visit |
|---|---|---|---|---|---|---|---|---|---|---|---|---|
| 1 | Male | A | Case | 40 | 187 | 80 | 5 | 37.3 | 3.8 | 88 | ··· | Baseline |
| 1 | Male | A | Case | 40 | 187 | 82 | 5 | NA | NA | 88 | ··· | 3 months |
| 1 | Male | A | Case | 40 | 187 | 81 | 5 | 37.4 | 3.8 | 89 | ··· | 6 months |
| 2 | Female | B | Control | 27 | 164 | 58 | 12 | 42.1 | 4.5 | 100 | ··· | Baseline |
| 2 | Female | B | Control | 28 | NA | NA | 12 | 42.0 | 4.3 | 100 | ··· | 3 months |
| 2 | Female | B | Control | 28 | 164 | 58 | 13 | 42.2 | 4.5 | 100 | ··· | 6 months |
| ⋮ | ⋮ | ⋮ | ⋮ | ⋮ | ⋮ | ⋮ | ⋮ | ⋮ | ⋮ | ⋮ | ⋱ | ⋮ |
| n | Female | B | Case | 56 | 175 | 68 | 10 | 38.7 | 3.6 | 60 | ··· | Baseline |
| n | Female | B | Case | 56 | 175 | 70 | 10 | 38.5 | 3.7 | NA | ··· | 3 months |
| n | Female | B | Case | 57 | 176 | 71 | 10 | 38.4 | 3.7 | 72 | ··· | 6 months |

(a)

**Unbalanced longitudinal data**

| Patient ID | Sex | Group | Age | ··· | Visit |
|---|---|---|---|---|---|
| 1 | Male | A | 26 | ··· | Baseline |
| 1 | Male | A | 26 | ··· | 12 days |
| 2 | Female | C | 32 | ··· | Baseline |
| 2 | Female | C | 32 | ··· | 13 days |
| 2 | Female | C | 32 | ··· | 17 days |
| 2 | Female | C | 32 | ··· | 21 days |
| ⋮ | ⋮ | ⋮ | ⋮ | ⋱ | ⋮ |
| n | Female | B | 29 | ··· | Baseline |
| n | Female | B | 29 | ··· | 12 days |
| n | Female | B | 29 | ··· | 24 days |

(b)

**Balanced longitudinal data**

| Patient ID | Sex | Group | Intervention | Age_B | Age_3 | Age_6 | Height_B | Height_3 | Height_6 | Weight_B | Weight_3 | Weight_6 | CCI_B | CCI_3 | CCI_6 | Albumin_B | ··· |
|---|---|---|---|---|---|---|---|---|---|---|---|---|---|---|---|---|---|
| 1 | Male | A | Case | 40 | 40 | 40 | 187 | 187 | 187 | 80 | 82 | 81 | 5 | 5 | 5 | 37.3 | ··· |
| 2 | Female | B | Control | 27 | 28 | 28 | 164 | NA | 164 | 58 | NA | 58 | 12 | 12 | 13 | 42.1 | ··· |
| ⋮ | ⋮ | ⋮ | ⋮ | ⋮ | ⋮ | ⋮ | ⋮ | ⋮ | ⋮ | ⋮ | ⋮ | ⋮ | ⋮ | ⋮ | ⋮ | ⋮ | ⋱ |
| n | Female | B | Case | 56 | 56 | 57 | 175 | 175 | 176 | 68 | 70 | 71 | 10 | 10 | 10 | 38.7 | ··· |

(c)

Figure 1: Illustration of key characteristics and different forms of longitudinal data. Subfigure (a) shows balanced (subjects have identical visit sequences), and subfigure (b) shows unbalanced (differing visit sequences) longitudinal data in long format (multiple rows per subject). The third subfigure (c) illustrates the same data as in (a) but in wide format (single row per subject). In this illustration, the focus could be in modeling the change in response variables such as quality of life (QoL) or albumin levels (reflecting disease progression) across intervention groups receiving distinct treatments, incorporating static variables (e.g., sex, group) and time-varying variables (e.g., age, height, weight) as covariates (features). The main difference to another common tabular data type, cross-sectional data, are precisely these repeatedly measured variables from the same subjects over time. The repeated measurements create a unique temporal structure that is essential for accurate analysis and synthetic data generation. Missing data (NA), measurement errors (176 in (a)), and dropouts (second row in (b)) are common issues encountered in longitudinal data and can complicate their analysis and synthetic data generation.

Longitudinal data are related but not equivalent to other time-dependent data types, of which time series and survival data are examples. Time series involve frequent, equally spaced repeated measurements, often with assumed dependencies between multiple series, and usually lack static variables [60]. Conversely, LD consist of independent realizations of relatively short subject-specific series of observations. The key difference between longitudinal and survival data lies in their conceptualization of time and its role in data analysis. In LD, time is treated as a fixed index, whereas in survival analysis, time-to-event is modeled as a random variable and is of primary interest [24].

### 1.1 Rationale

Both regular tabular, i.e., cross-sectional, and longitudinal SD generators encounter challenges when facing different variable types or missing data. However, using a regular tabular data generator for LD can result in flawed generative models and manifest as logical inconsistencies, such as treating repeated within-subject measurements as independent entities, leading to eventual loss of temporal correlation [82], or inaccurately representing variability within or between subjects [7]. Cross-sectional data generators also struggle with the sparsity of unbalanced LD and cannot separate "true missingness" from the unbalanced structure. While synthetic time-series generators can capture temporal relationships, they may inadequately replicate the underlying correlation structures due to their reliance on large volumes of equispaced repeated measurements. Furthermore, these generators often struggle to produce static variables in conjunction with the time series. Synthetic survival data generators focus on modeling a single time-to-event as a continuous variable, lacking potential in generating repeated measurements at discrete time points in LD.

Consequently, additional research is warranted to identify appropriate techniques for generating synthetic longitudinal patient data that are reliable and of sufficient quality to be used in real-life settings. Such methods could be directly offered to data controllers to facilitate the use of patient data in different RDI and educational activities while safeguarding patient privacy. This systematic literature review aims to address this need and serves as a natural continuation to a number of previous reviews on SDG methods in healthcare [19, 20, 32, 43, 47] and in the broader context of tabular data [39]. While almost all preceding reviews recognize the prevalence of longitudinal data, none of them focus on those. Rather, in all these reviews the treatment of LD with respect to both methods and their evaluation is largely overshadowed by cross-sectional data. Some reviews (e.g. [20]) discuss the inherent characteristics of longitudinal data such as balance, yet these and other crucial aspects of LD (see Figure 1) are not collected from the reviewed methods. One study [45] presents a list of SDG methods for LD as motivation for their methodological research but provides no further discussion about the methods' properties or their evaluation.

Our systematic review addresses a significant gap in the literature by comprehensively surveying a critical type of medical data that has been acknowledged for their importance but never thoroughly reviewed in the context of SDG. Adhering to PRISMA guidelines [51] throughout our review distinguishes our work from previous studies, which often offered only partial compliance, typically limited to a flow diagram (see Figure 2). This adherence not only demonstrates the feasibility of conducting a systematic methodological review according to established quality standards but also emphasizes the rigor of our approach. Moreover, unlike earlier reviews, we frame our analysis entirely through the lens of LD, highlighting often-overlooked aspects such as the evaluation of temporal dependencies and the preservation of statistical inferences—topics that have not been adequately addressed in prior literature. Our review encompasses studies published up to May 2024, capturing recent advancements in generative artificial intelligence. While it is challenging to cover every emerging SDG approach due to the rapid pace of innovation, our results indicate that the majority of methods have emerged in the last two years, underscoring the relevance and timeliness of our findings.

### 1.2 Objectives

The primary objective of this systematic review is to map and describe existing methods of generating synthetic longitudinal patient data in real-life settings. The research questions under the primary objective are:

**RQ1.** What methods are currently available for generating synthetic longitudinal patient data?

**RQ2.** How do these methods address the key characteristics of longitudinal data, including temporal structure, balance, different variable types, and missing values?

**RQ3.** How were these methods evaluated in terms of resemblance, utility, and privacy preservation?

The secondary objectives are to evaluate the comprehensiveness of reporting in the identified literature and to provide insights for method developers about areas requiring further research. Comparing the identified methods in practice is beyond the scope of this review but presents an intriguing prospect for future research.

The rest of the article is organized as follows: Section 2 outlines the review's methodology, Section 3 presents the findings of individual studies and their synthesis. Section 4 concludes the article by offering general interpretations of the results, addressing limitations and discussing practical implications.

# 2 MATERIALS AND METHODS

## 2.1 Eligibility criteria

To address RQ1–3, we established specific selection criteria outlined in Table 1. We delineated the concept of synthetic data, acknowledging the challenge of distinguishing between simulated and synthetic data in the existing literature (criterion 1). In consideration of real-world applicability, we excluded overly simplistic methods where SD are generated from standard probability distributions (such as multivariate normal) after estimating their parameters, as well as theoretical simulation models deemed too basic to capture real-world complexities.

Table 1: Eligibility criteria. The following criteria were applied to select relevant literature from the search results using the specific screening charts presented in Appendix A.2.

| Number | Criterion | Description | Examples of exclusions |
|---|---|---|---|
| 1 | Includes synthetic data generation | Synthetic data are generated via a randomized algorithm using an existing dataset (i.e., original or input dataset) with the goal of closely mimicking the original data distribution and with the ability to generate unlimited number of synthetic samples. | • Deterministic algorithm, e.g., rule-based<br>• Algorithm based naively on standard probability distributions<br>• Data simulation, i.e., data are generated from theoretical models |
| 2 | Longitudinal data | The input and output dataset includes at least one repeatedly measured variable, and the authors address this longitudinal aspect. | • Not longitudinal data<br>• Longitudinal data altered so that the temporal structure is lost, e.g., through aggregation<br>• Variables included "incidentally" without explicitly considering repeated measurements or temporal correlation |
| 3 | Data comparability | In case the input data are not patient data, variables in the original dataset should be comparable to those found generally in longitudinal patient data. | • Data not comparable to longitudinal patient data |
| 4 | Data sharing capability | Method should support data sharing and generate fully synthetic data which is void of the original confidential data. | • Publication generates partially synthetic data |
| 5 | Privacy-preserving techniques | Consideration for privacy-preserving techniques in SDG is optional. | |
| 6 | Non-open-source and commercial methods | Literature involving non-open-source and commercially licensed methods are included. | |
| 7 | Language | English | • Language other than English |
| 8 | Publication types | Peer-reviewed journals and proceedings as well as pre-prints, books, book chapters and reviews. | • Other than listed in the Description |

Given our focus on longitudinal patient data, compatibility with this data type was essential (criteria 2–3). With the overarching goal of facilitating data sharing in healthcare, we sought methods capable of generating fully synthetic individual-level data (criterion 4). Consequently, publications generating partially synthetic data, either by generating only a subset of variables or specific types of observations to mitigate class imbalances, were excluded.

Privacy considerations are paramount in data sharing. However, due to the lack of official guidelines and consensus on additional privacy measures, we took a permissive approach, not mandating separate privacy-preserving mechanisms (criterion 5). Wanting to provide a wide range of methods for different RDI activities, we did not limit our scope to open-source methods and included non-peer-reviewed literature, while restricting our selection to English (criteria 6–8).

In secondary use of health data, LD are commonly derived from EHRs stored in a hospital database. EHRs integrate tabular data, specifically individual-level microdata, consisting of different patient characteristics and medical events alongside other data types such as imaging, signal and text data, making them multimodal. As the focus in this review is on LD, methods generating EHRs were included if their tabular data component met criteria 2–3.

### 2.2 Information sources

We searched EMBASE (1947/01/01—2024/05/22), MEDLINE (Ovid interface, 1946/01/01—2024/05/22), Web of Science (1900/01/01—2024/05/22) and Google Scholar (Publish or Perish software [29], first 1000 hits on 2021/06/18), and arXiv (open-source metadata [11] on 2024/05/22). These databases were chosen because they offer the best coverage [8] and the latest, unpublished methods.

### 2.3 Search strategy

Search algorithms were developed using topic (title, abstract, keywords) and text words related to synthetic longitudinal patient data. The search algorithms, detailed in Appendix A.1, utilized terms such as 'synthetic', 'artificial', 'data' and 'record', intended to cover all possible synonyms used to refer to synthetic data in the literature. Words such as 'longitudinal', 'follow-up', 'trajectory', 'health', 'medical' and 'patient' were used to capture works that deal with longitudinal patient data. The initial algorithm was reviewed by an independent review team member using the PRESS standard [44]. To ensure timeliness, the search was conducted thrice in June 2021, November 2022 and May 2024. Additional records were identified through manual screening of the references in the eligible publications.

### 2.4 Selection process

After removing duplicates using EndNote Online [67], the results were uploaded to Rayyan [49] for literature screening. During the first two search iterations, the review authors (KP and JV) independently screened the titles and abstracts against the eligibility criteria (Table 1) using a specific screening chart (Appendix A.2.1) and leveraged Rayyan's features to emphasize keywords indicative of inclusion or exclusion, streamlining the selection process. Any remaining duplicates were removed by KP. Subsequently, full texts, referred to as studies or publications, were procured for records meeting the eligibility criteria or exhibiting any uncertainty in eligibility. Inaccessible publications were excluded. KP and JV independently assessed these full texts against the eligibility criteria using a specific full-text screening chart (Appendix A.2.2). Disagreements were resolved through discussion and, if necessary, a third-party arbitration (KA) was consulted. The reasons for exclusion were documented. The third search iteration followed the same principles, with KP conducting the screening and consulting JV for uncertain cases.

### 2.5 Data items and collection

KP collected and managed data from the eligible publications using a structured form (Appendix A.3) within the REDCap electronic data capturing tool [27, 28]. In unclear situations, KP consulted the authors and their webpages to make sure that all relevant data were captured. For quality control, JV and KA conducted spot checks on the data collection process. Unavailable or unclear information was recorded as missing, and in cases of uncertainty, KP conferred with JV and KA.

### 2.6 Risk of bias and reporting quality assessment

In adherence to the PRISMA guidelines [51], a systematic review is expected to consider and assess potential biases present in the included studies. Due to the absence of a specific evaluation framework in our context of a methodological review, we applied the existing guidelines from the Cochrane Handbook for Systematic Reviews of Interventions [7]. We considered events such as favoring data subsets for better results (selection bias), omission of results (reporting bias), and ambiguity in model performance assessment (performance bias). The detailed assessment framework is presented in Appendix A.4.1.

In the assessment of reporting quality, we adopted criteria inspired by the Grading of Recommendations, Assessment, Development and Evaluations framework [25], focusing on aspects such as the use of identical terminology or notations to denote disparate phenomena without clarification (inconsistency), a lack of precision or clarity in presenting information (imprecision), or conveying information in a less straightforward or explicit manner (indirectness), albeit to a lesser degree than in reporting bias. These factors pose a risk of misinterpretation, potentially leading to ambiguity or comprehension difficulties. Further details are provided in Appendix A.4.3.

KP assessed the risk of bias and the reporting quality, sharing the findings with JV and KA to obtain mutual agreement. Furthermore, KP collected data related to disclosed conflicts of interest and peer-review status.

## 2.7 Methods to address research questions

To address RQ1, we compiled a concise overview encompassing all SDG methods that were predominantly applied in the eligible publications, subsequently referred to as the primary methods (Section 3.4). In addition, we recorded all SDG methods that were utilized for comparing and benchmarking against the primary methods, later referred to as the reference methods.

To address RQ2, we constructed a comprehensive table (Section 3.4) outlining the requirements, characteristics, and capabilities of the primary methods. Recognizing the diversity of longitudinal datasets found in the literature, we grouped them into three categories. Methods generating data with both static and time-varying variables were classified as producing “standard data”. Those generating only time-varying variables without static variables were classified as producing “trajectory data” (e.g., diagnostic codes, procedure codes, and prescriptions per visit). Finally, methods generating a single repeatedly measured variable, such as a disease code, were classified as producing “sequence data”.

To address RQ3, we constructed summary figures and tables outlining the utilized datasets and means to evaluate resemblance, utility and privacy (Section 3.5). To gain insight into the broader evaluation framework, we examined whether each evaluation task was conducted using a single or multiple independently generated synthetic datasets and whether the evaluation of the SD quality was conducted in relation to the original data or some other way.

In accordance with the secondary objectives, we categorized the research objectives of the eligible publications (Section 3.2) , classified the primary methods based on their main operating principle (Section 3.4) and assessed bias and reporting quality (Section 3.3). Finally, we discussed our findings in relation to the existing literature to identify future research areas (Section 4).

# 3 RESULTS

## 3.1 Study selection

The search initially identified 11 307 publications. After removing 2 027 duplicates, 8 605 studies underwent title and abstract screening, leading to selection of 517 publications for full-text screening. Nine studies were unattainable, leaving 508 studies for evaluation against the eligibility criteria (Section 2.1). Altogether 33 of the 508 studies met our criteria and were included at this stage. To augment the search, KP examined all references in the 33 included studies. This process identified 24 potential publications, of which three were deemed eligible and incorporated in the review. Ultimately, 36 eligible studies were included in the review. Figure 2 illustrates the study selection process according to the PRISMA guidelines [51].

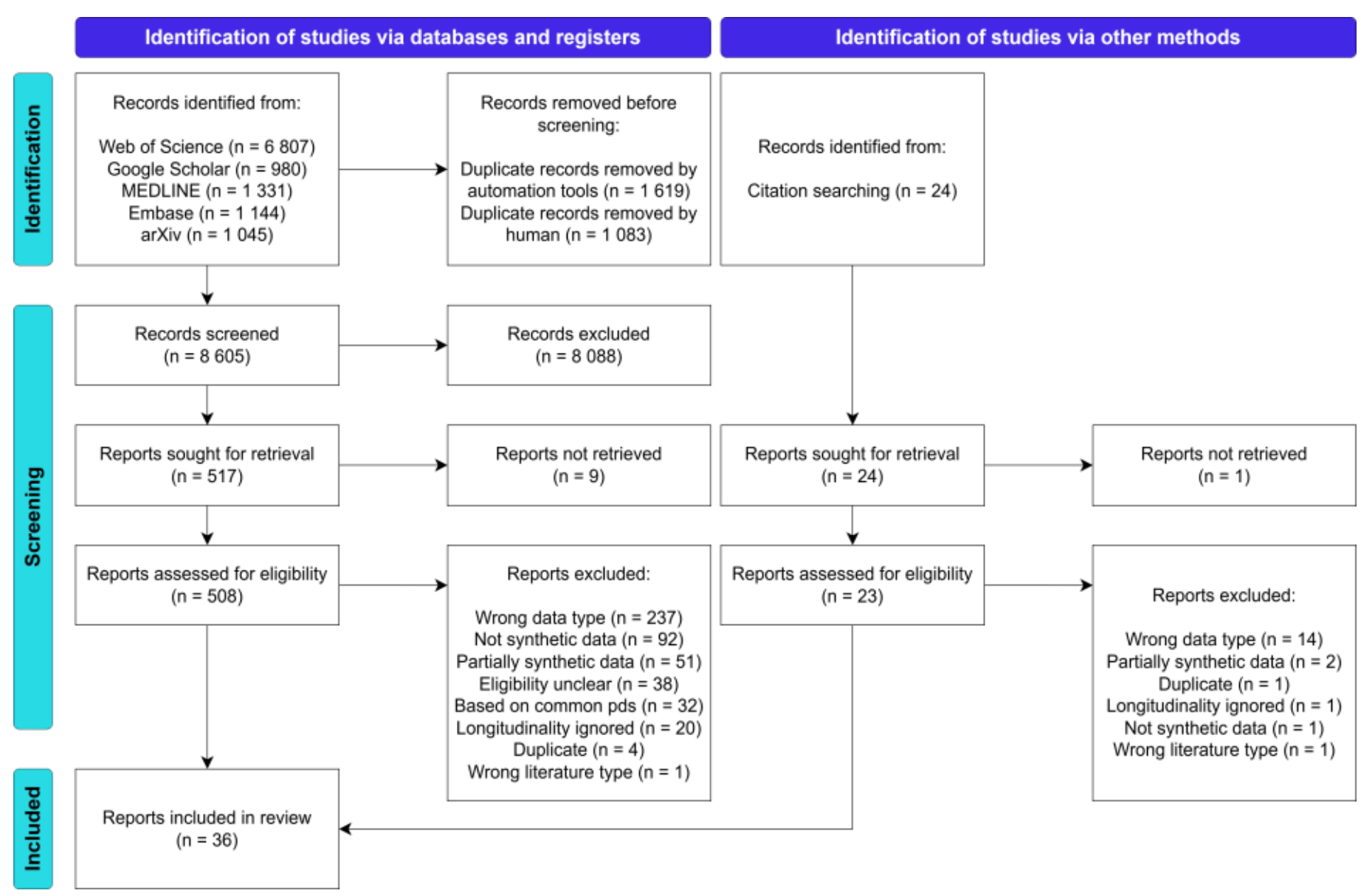

Figure 2: PRISMA flow diagram. The diagram illustrates the study selection process according to the PRISMA guidelines; pds: probability distributions. Examples of each exclusion category are provided in Appendix A.5.

## 3.2 Study characteristics

The 36 included studies (Table 2) were published between 2016–2024, with 13 (36%) published in 2023 and nine (25%) in 2022. The predominant research objective in the studies was privacy-preserving data publishing (64%). Additionally, eight (22%) studies emphasized data publishing but did not employ privacy-preserving techniques or privacy evaluation. As per SCImagoJR [58], the most common publication fields were computer science (50%) and medicine (33%).

Table 2: Summary of the included publications. The table provides a summary of the 36 publications included in the systematic literature review. The publications are presented in descending order of publication year and sorted alphabetically by author name. The table includes information about the type of publication, the field based on SCImago Journal & Country Rank [58], and the objective of each publication as interpreted by the review authors.

| Authors | Journal | Type | Field | Objective | Year |
|---|---|---|---|---|---|
| Belgodere et al. [4] | arXiv | Preprint | Multidisciplinary | Framework development | 2024 |
| Bun et al. [9] | Proceedings of the ACM on Management of Data | Journal article | Computer science | PPDP | 2024 |
| Kühnel et al. [35] | Scientific Reports | Journal article | Multidisciplinary | (PP)DP | 2024 |
| Pang et al. [52] | arXiv | Preprint | Multidisciplinary | PPDP | 2024 |
| Theodorou et al. [69] | The Thirty-Eighth AAAI Conference on Artificial Intelligence | Conference paper | Computer science | (PP)DP | 2024 |
| Das, Wang & Sun [12] | Proceedings of the ACM SIGKDD International Conference on Knowledge Discovery and Data Mining | Conference paper | Computer science | PPDP | 2023 |
| El Kababji et al. [34] | JCO clinical cancer informatics | Journal article | Biochemistry, genetics and molecular biology<br>Medicine | PPDP | 2023 |
| Haleem et al. [26] | Electronics (Switzerland) | Journal article | Computer science<br>Engineering | (PP)DP | 2023 |
| Hashemi et al. [30] | Proceedings of the International Joint Conference on Neural Networks | Conference paper | Computer science | PPDP | 2023 |
| Kuo et al. [36] | Journal of Biomedical Informatics | Journal article | Computer science<br>Medicine | PPDP | 2023 |
| Kuo et al. [37] | NeurIPS 2023 Workshop on Synthetic Data Generation with Generative AI | Conference paper | Computer science | PPDP | 2023 |
| Li et al. [41] | npj Digital Medicine | Journal article | Computer science<br>Health professions<br>Medicine | PPDP | 2023 |
| Mosquera et al. [45] | BMC Medical Research Methodology | Journal article | Medicine | PPDP | 2023 |
| Nikolentzos et al. [48] | npj Digital Medicine | Journal article | Computer science<br>Health professions<br>Medicine | PPDP | 2023 |
| Sood et al. [63] | PLOS ONE | Journal article | Multidisciplinary | PPDP | 2023 |
| Sun, Lin & Yan [66] | arXiv | Preprint | Multidisciplinary | PPDP | 2023 |
| Theodorou, Xiao & Sun [70] | Nature Communications | Journal article | Biochemistry, genetics and molecular biology<br>Chemistry<br>Physics and Astronomy | PPDP | 2023 |
| Yoon et al. [79] | npj Digital Medicine | Journal article | Computer science<br>Health professions<br>Medicine | PPDP | 2023 |
| Bhanot et al. [5] | Neurocomputing | Journal article | Computer science<br>Neuroscience | Resemblance quantification | 2022 |
| Kuo et al. [38] | Scientific Data | Journal article | Computer science<br>Decision science<br>Mathematics<br>Social Sciences | PPDP | 2022 |
| Lu et al. [42] | arXiv | Preprint | Multidisciplinary | (PP)DP | 2022 |
| Shi et al. [59] | Frontiers in Artificial Intelligence | Journal article | Computer science | PPDP | 2022 |
| Wang & Sun [75] | Proceedings of the Conference on Empirical Methods in Natural Language Processing | Conference paper | Computer science | PPDP | 2022 |

| Authors | Journal | Type | Field | Objective | Year |
|---|---|---|---|---|---|
| Wang et al. [74] | BMC Medical Informatics and Decision Making | Journal article | Computer science<br>Medicine | Modeling | 2022 |
| Wendland et al. [76] | npj Digital Medicine | Journal article | Computer science<br>Health professions<br>Medicine | (PP)DP | 2022 |
| Yu, He & Raghunathan [80] | Journal of Survey Statistics and Methodology | Journal article | Decision sciences<br>Mathematics<br>Social sciences | PPDP | 2022 |
| Zhang, Yan & Malin [82] | Journal of the American Medical Informatics Association | Journal article | Medicine | Framework development | 2022 |
| Biswal et al. [6] | Proceedings of Machine Learning Research: Machine Learning for Healthcare | Conference paper | Computer science and technology<br>Computing<br>Data processing | PPDP | 2021 |
| Zhang et al. [81] | Journal of the American Medical Informatics Association | Journal article | Medicine | PPDP | 2021 |
| Gootjes-Dreesbach et al. [23] | Frontiers in Big Data | Journal article | Computer science | PPDP | 2020 |
| Sood et al. [62] | Scientific Reports | Journal article | Multidisciplinary | (PP)DP | 2020 |
| Beaulieu-Jones et al. [3] | Circulation: Cardiovascular Quality and Outcomes | Journal article | Medicine | PPDP | 2019 |
| Fisher et al. [18] | Scientific Reports | Journal article | Multidisciplinary | Modeling | 2019 |
| Barrientos et al. [2] | The Annals of Applied Statistics | Journal article | Decision sciences<br>Mathematics | PPDP | 2018 |
| Walonoski et al. [72] | Journal of the American Medical Informatics Association | Journal article | Medicine | (PP)DP | 2018 |
| Raab, Nowok & Dibben [54] | Journal of Privacy and Confidentiality | Journal article | Computer science<br>Mathematics | (PP)DP | 2016 |

PPDP: privacy-preserving data publishing; (PP)DP: data publishing without considering data privacy.

### 3.3 Risk of bias and reporting quality

The included studies showed no indication of selection bias. However, 11 studies (28%) had potential risk of performance bias due to inadequate transparency in their evaluation descriptions, particularly regarding model training of reference methods, or due to a lack of source code availability to verify the evaluation processes. A risk of reporting bias was observed in 17 publications (44%). While the specific issues varied, the primary reason was the absence or partial presentation of results that were described in the methodology but not reported in the main results or supplemental materials. Further details are available in Appendix A.4.2.

Inconsistency of reporting was observed in one study, imprecision in six (15%), and evidence of indirect reporting in seven studies (18%). These findings related to presentation of p-values, sample size, included variables, evaluation metrics and privacy budget with differing degrees of precision or notation. Further details are given in Appendix A.4.3.

In five studies (13%) the documentation of the training processes was only partially provided, and 15 publications (38%) lacked them altogether. Among the included publications, only four (11%) were not peer reviewed, consisting exclusively of arXiv preprints. Conflict of interest was declared in eight (21%) publications, whereas this information was absent in 14 (36%) publications.

### 3.4 SDG methods for longitudinal patient data

We identified a total of 66 SDG methods, comprising 39 primary and 27 reference methods (Appendix A.6). Figure 3 illustrates a categorization of all identified methods according to their key operating principles while Table 3 details the key characteristics of the primary methods. Most methods (67%) were designed to generate standard LD, while eight methods (21%) generated sequence data and six (15%) generated trajectories. With the exception of Synthea, which uses a pre-built disease module generator but allows user-created modules [68], all primary methods permitted user customization or training for specific datasets. However, 10 methods (26%) required expert knowledge, meaning they demand detailed, data-specific information for effective use.

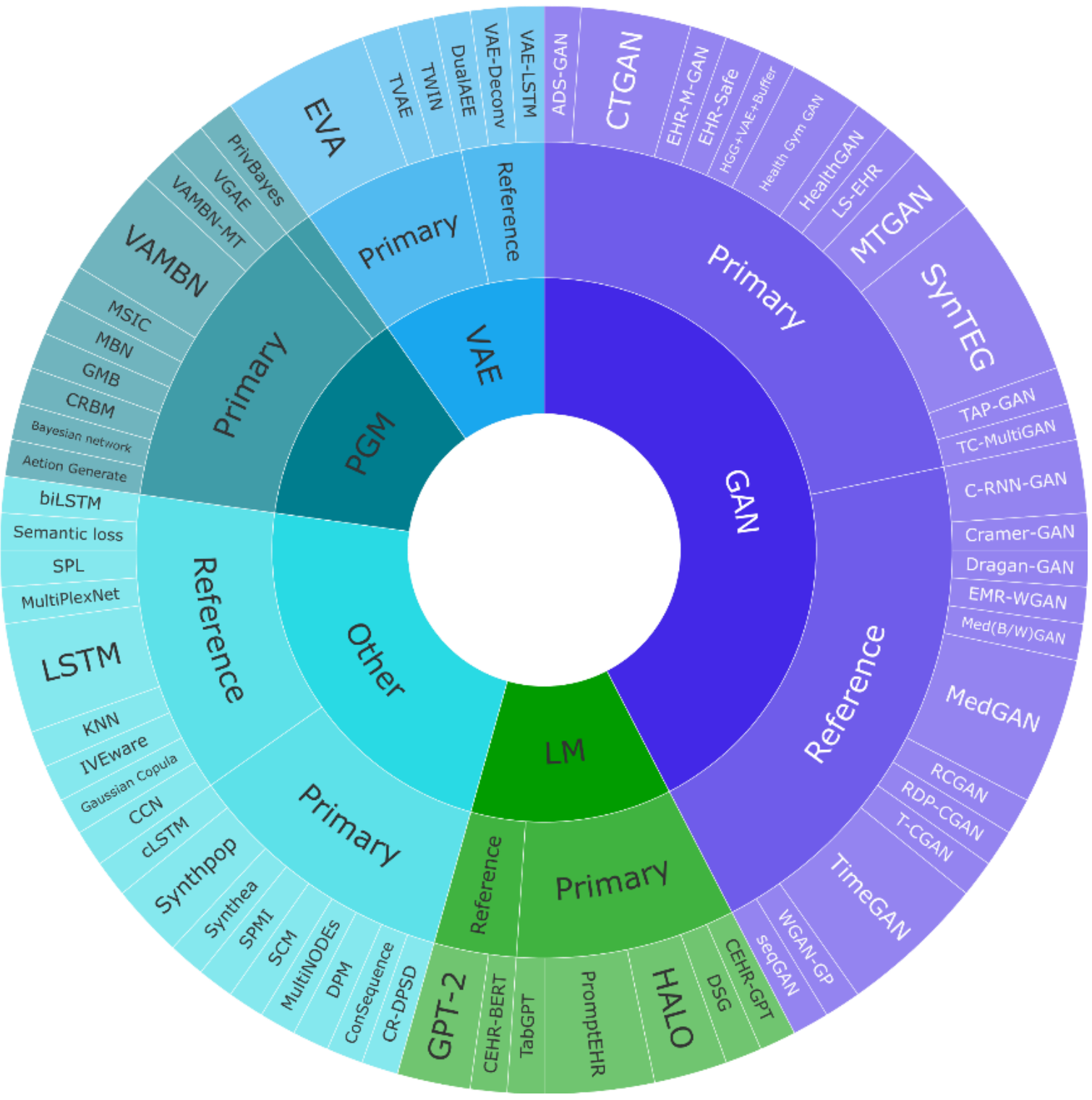

Figure 3: Categorization of the identified SDG methods. The 39 primary methods and 27 reference methods identified were classified into five distinct groups according to their main operating principle: generative adversarial networks (GANs), language models (LMs), variational autoencoders (VAEs), probabilistic graphical models (PGMs), and others. The width of each sector represents the frequency of method use across the 36 publications reviewed.

Different method categories employed varying approaches to model the temporal structure of LD. Generative adversarial networks (GANs) utilized convolutional layers and recurrent neural networks (RNNs), such as gated recurrent units or long short-term memory (LSTM) networks, which are adept at handling sequential data [10, 21, 22, 57]. Probabilistic graphical models (PGMs) captured temporal structure by constraining node interactions, ensuring future values being modeled based on preceding values and relevant covariates. Variational autoencoders (VAEs) utilized autoregressive techniques, feeding data sequentially while latent variables accounted for patient-specific heterogeneity. Language models (LMs) modeled temporal data in a manner similar to sequential text generation utilizing transformer architectures and self-attention mechanisms [71]. Other methods employed a variety of techniques, but the most prevalent approaches were (a) the Markov assumption, where the present depends only on the previous time point, and (b) ordered generation of each subsequent variable, including the non-temporal variables, conditionally on all previously generated variables (instead of just the most recent one).

Table 3: Summary of the 39 identified primary synthetic longitudinal data generation methods. The table summarizes each method, its classification, the type of longitudinal synthetic data generated, its ability to generate unbalanced and missing data structures, as well as categorical and continuous variables (✓ Yes, ✗ No, ? Unclear). Additionally, the table specifies whether expert knowledge is required, the availability of source code and pseudocode (referenced in parentheses if not from the listed publication), and the programming language used. The final column lists all publications where the method was applied. Ideally, the optimal method would have a ✓ in every column

| Method | Class | Data type | Unbalanced data | Missing data | Categorical variables | Numerical variables | No expert knowledge | Privacy mechanism | Source / pseudocode | Prog. language | Applied in |
|---|---|---|---|---|---|---|---|---|---|---|---|
| AC-GAN | GAN | Standard | ? | ✗ | ✓ | ✓ | ✓ | ✓ | ✓ / ✗ | Python | [3] |
| ADS-GAN | GAN | Standard | ? | ✗* | ✓ | ✓ | ✓ | ✓ | ✓ [53] / ✗ | Python | [59] |
| CTGAN | GAN | Standard | ✓ | ✗ | ✓ | ✓ | ✓ | ✗ | ✓ [53] / ✗ | Python | [26, 34, 42] |
| EHR-Safe | GAN | Standard | ✓ | ✓ | ✓ | ✓ | ✓ | ✗ | ✗ / ✓ | Python | [79] |
| Health Gym GAN | GAN | Standard | ? | ✗ | ✓ | ✓ | ✓ | ✗ | ✓ / ✗ | Python | [36, 38] |
| HealthGAN | GAN | Standard | ? | ✗ | ✓ | ✓ | ✓ | ✗ | ✓ [77] / ✗ | Python | [5] |
| HGG-VAE-Buffer | GAN | Standard | ✗ | ✗ | ✓ | ✓ | ✓ | ✗ | ✗ / ✓ | Python | [36] |

| Method | Class | Data type | Unbalanced data | Missing data | Categorical variables | Numerical variables | No expert knowledge | Privacy mechanism | Source / pseudocode | Prog. language | Applied in |
|---|---|---|---|---|---|---|---|---|---|---|---|
| TC-MultiGAN | GAN | Standard | ✗ | ✗* | ✓ | ✓ | ✓ | ✗ | ✗ / ✗ | Python | [26] |
| CEHR-GPT | LM | Standard | ✓ | ? | ✓ | ✗ | ✓ | ✗ | ✗ / ✗ | ? | [52] |
| DSG | LM | Standard | ✓ | ✓ | ✓ | ✓ | ✓ | ✗ | ✗ / ✗ | Python | [26] |
| TabGPT | LM | Standard | ? | ? | ✓ | ✓ | ✓ | ✓ | ✓ [50] / ✗ | Python | [4] |
| HALO | LM | Standard<br>Sequence | ✓ | ✓ | ✓ | ✓ | ✓ | ✗ | ✓ / ✗ | Python | [69, 70] |
| Aetion Generate† | PGM | Standard | ✗ | ? | ✓ | ✓ | ✓ | ? | ✗ / ✓ [16] | Python + R | [34] |
| Bayesian network | PGM | Standard | ? | ✗* | ✓ | ✓ | ✗ | ✗ | ✗ / ✗ | R | [63] |
| MBN | PGM | Standard | ✗ | ✗* | ✓ | ✓ | ✗ | ✗ | ✓** / ✗ | R | [62] |
| VAMBN | PGM | Standard | ? | ✗* | ✓ | ✓ | ✗ | ✗ | ✓ / ✗ | Python + R | [23, 35, 76] |
| VAMBN-MT | PGM | Standard | ? | ✗* | ✓ | ✓ | ✗ | ✗ | ✓ / ✗ | Python + R | [35] |
| TVAE | VAE | Standard | ✓ | ✗ | ✓ | ✓ | ✓ | ✗ | ✓ [53] / ✗ | Python | [34] |
| cLSTM | Other | Standard | ✓ | ✓ | ✓ | ✓ | ✓ | ✗ | ✓** / ✗ | Python | [45] |
| CRBM | Other | Standard | ✗ | ✗* | ✓ | ✓ | ✓ | ✗ | ✗ / ✗ | ? | [18] |
| DPM | Other | Standard | ? | ✗ | ✓ | ✓ | ✓ | ✗ | ✗ / ✗ | Python | [37] |
| MultiNODEs | Other | Standard | ? | ✗* | ✓ | ✓ | ✓ | ✗ | ✓ / ✗ | Python + R | [76] |
| SCM | Other | Standard | ? | ✓ | ✓ | ✓ | ✗ | ✓ | ✓ / ✗ | R | [2] |
| SPMI | Other | Standard | ? | ✗* | ✓ | ✓ | ✓ | ✗ | ✓** / ✗ | R | [80] |
| Synthea | Other | Standard | ? | ✗ | ✓ | ✓ | ✗ | ✗ | ✓ / ✗ | Java | [72] |
| Synthpop | Other | Standard | ? | ✓ | ✓ | ✓ | ✗ | ✗ | ✓ / ✗ | R | [54, 80] |
| EHR-M-GAN | GAN | Trajectory | ? | ✗ | ✓ | ✓ | ✓ | ✓ | ✓ / ✓ | Python | [41] |
| TAP-GAN | GAN | Trajectory | ? | ✗ | ✓ | ✓ | ✓ | ✓ | ✗ / ✗ | Python | [30] |
| PromptEHR | LM | Trajectory | ? | ? | ✓ | ✗ | ✓ | ✗ | ✓ / ✗ | Python | [12, 66, 75] |
| MSIC | PGM | Trajectory | ? | ? | ✓ | ✗ | ✓ | ✗ | ✗ / ✓ | ? | [66] |
| VGAE | PGM | Trajectory | ? | ? | ✓ | ? | ? | ✓ | ✓** / ✗ | ? | [48] |
| TWIN | VAE | Trajectory | ✗ | ? | ✓ | ✗ | ✓ | ✗ | ✓ / ✗ | Python | [12] |
| LS-EHR | GAN | Sequence | ? | ✗ | ✓ | ✗ | ✓ | ✗ | ✗ / ✗ | ? | [82] |
| MTGAN | GAN | Sequence | ? | ✗ | ✓ | ✗ | ✓ | ✗ | ✓ / ✓ | Python | [42, 66] |
| SynTEG | GAN | Sequence | ? | ✗ | ✓ | ✗ | ✓ | ✗ | ✗ / ✗ | ? | [12, 41, 70, 75, 81] |
| GMB | PGM | Sequence | ✓ | ✗ | ✓ | ✗ | ✗ | ✗ | ✗ / ✓ [73] | Python | [74] |
| EVA | VAE | Sequence | ✓ | ✗ | ✓ | ✗ | ✓ | ✗ | ✗ / ✗ | ? | [6, 12, 66, 70] |
| ConSequence | Other | Sequence | ✓ | ✓ | ✓ | ✗ | ✗ | ✗ | ✓ / ✗ | Python | [69] |
| CR-DPSD | Other | Sequence | ✗ | ✗ | ✓ | ✗ | ✗ | ✓ | ✗ / ✓ | ? | [9] |

Information about the method's ability to generate unbalanced data was available for 17 primary methods (44%), of which 10 were capable of performing this task, though three were restricted to sequence data. Seven primary methods (18%) could generate missing observations, although this often resulted from generating unbalanced LD rather than from a dedicated mechanism designed to replicate missing value patterns. Nine methods (23%) required imputation of missing values in the original data before generating SD. While all primary methods were capable of generating categorical variables, many required one-hot encoding. Out of the 39 primary methods, 27 (69%) were able to generate numerical variables; however, some of these methods required discretization of numerical data before SDG. Privacy mechanisms were incorporated in eight methods (21%), including four utilizing differential privacy (DP) [1, 15], two incorporating penalties in the optimization algorithm, and two applying other noise perturbation.

Source codes were available for 23 primary methods (59%), with five (13%) available in another publication and four upon request. Pseudocodes were given for eight methods (21%), two of which were provided in a cited publication. Both the source and pseudocodes were inaccessible for 10 methods (26%). Python was the most common programming language (64%), followed by R (23%), and the programming language for eight primary methods (21%) was unverifiable. System requirements were detailed in 11 studies (28%).

### 3.5 Evaluation approaches for synthetic data quality

The majority of studies (69%) generated a single synthetic dataset, while 25% produced a small number (<50) of datasets, and 2 studies (6%) generated a larger quantity (≥50). Twelve studies (33%) explored the impact of adjusting hyper parameters or altering the method's structural configuration on the quality of synthetic data, and 16 studies (44%) compared the primary method to reference methods (see Appendix A.6). The MIMIC-III dataset [33] was the most frequently used, appearing in 11 studies (31%), with a complete list of datasets available in Appendix A.7. Two studies augmented their assessment with simulated data.

Almost all studies (94%) compared SD to the original data in at least one of the following areas: resemblance, utility, or privacy, with [4] performing these comparisons in the embedding space. The remaining two studies pursued alternative approaches: [72] compared prevalence statistics in SD against empirical population data, and [74] focused solely on describing the characteristics of the SD. The subsequent subsections detail the approaches used to evaluate resemblance, utility, and privacy.

### *3.5.1 Resemblance*

Thirty-three studies (92%) assessed resemblance between the synthetic and original data. We discerned four different domains of resemblance: the similarity of univariate distributions, bivariate distributions, multivariate distributions, and temporal structure. Within each domain, we categorized the approaches into four paradigms: qualitative, quantitative, model-based, and statistical test-based (Figure 4). Specific approaches that did not align with these commonly used approaches were classified as "Other".

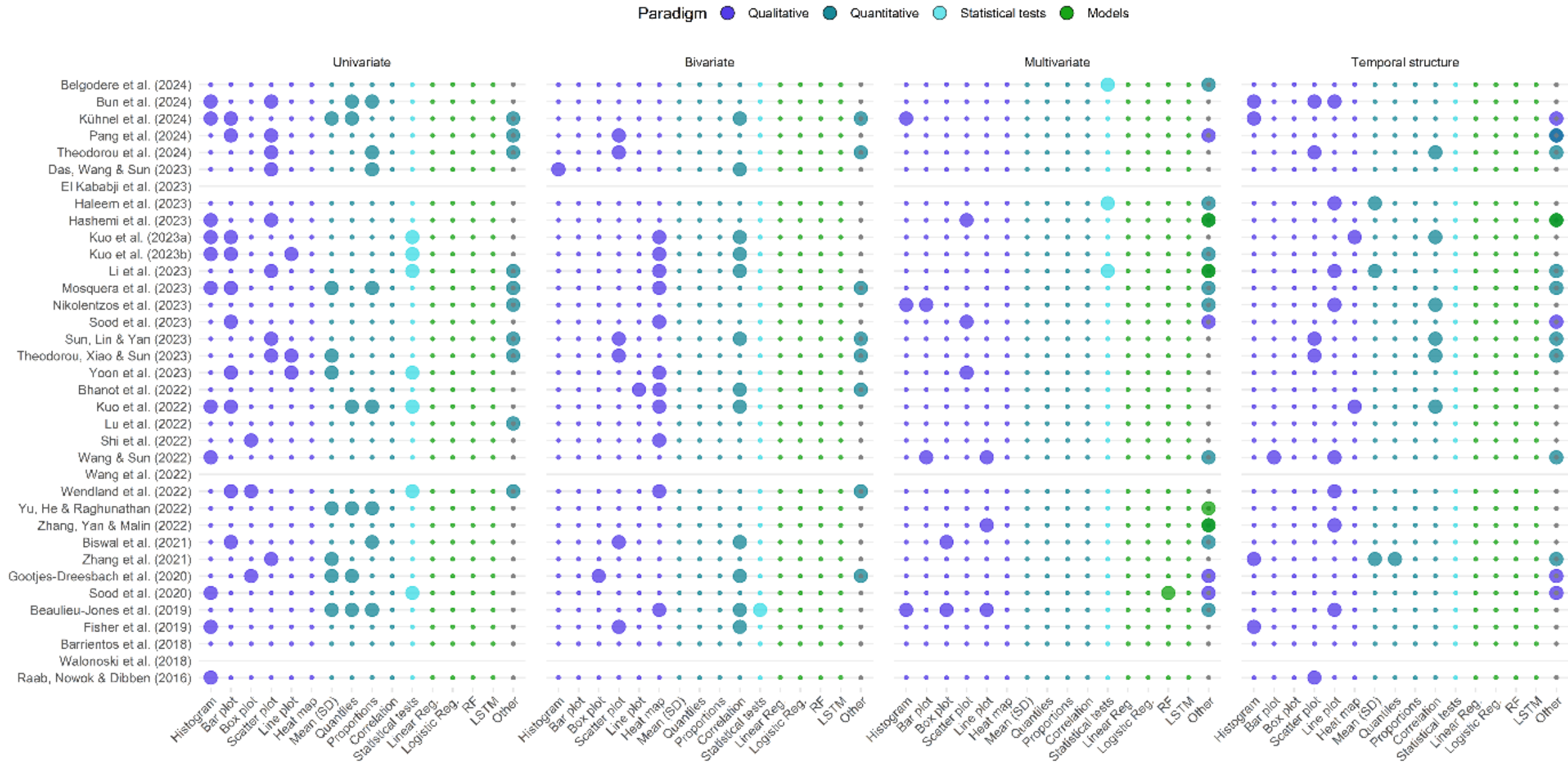

Figure 4: Approaches used to evaluate synthetic and original data resemblance. The figure illustrates the application of various approaches (shown on the x-axis) for assessing the resemblance between the synthetic and original data in the included studies, with the larger circles indicating the approach was used. The evaluation of resemblance is categorized into four domains: univariate distributions, bivariate distributions, multivariate distributions, and temporal structure. These approaches are further classified into qualitative, quantitative, model-based, and statistical test-based paradigms. It is important to note that a single study may employ multiple assessments within the same approach. The "Histogram" approach also includes density plots, and the "Box plot" approach encompasses violin plots. Statistical tests and approaches labeled as "Other" are discussed in the main text. SD: standard deviation; Reg.: regression; RF: random forests; LSTM: long short-term memory.

Among the 33 studies, seven (21%) evaluated all four domains. Similarly, 14 studies (42%) evaluated three domains, eight studies evaluated two domains, and three studies (9%) evaluated one domain. Univariate resemblance was assessed in 28 studies (85%), 20 studies (61%) examined bivariate resemblance, 18 studies (55%) evaluated multivariate resemblance, and 23 studies (70%) investigated temporal resemblance.

Univariate resemblance (Figure 4, panel 1) involved comparing univariate distributions between the synthetic and original data, i.e., comparing the distributions of one variable at a time. Qualitative paradigms were applied in 24 studies (73%), histograms being the most prevalent approach. Quantitative paradigms were employed in 19 studies (58%), primarily focusing on evaluating means, standard deviations, quantiles, and proportions. "Other" approaches to quantitative comparisons included, e.g., root mean square error (RMSE) and Jensen-Shannon divergence. Statistical tests were employed in seven studies (21%), including Mann-Whitney U, t-, or Kolmogorov-Smirnov (KS) tests for continuous variables, and Chi-squared homogeneity test for categorical variables.

Bivariate resemblance (Figure 4, panel 2) focused on contrasting the associations between pairs of variables in the synthetic data versus those in the original data. Quantitative and qualitative paradigms were used in 19 (58%) and 16 (48%) studies, respectively. The most common approaches were comparing correlations and the corresponding heatmaps. "Other" quantitative approaches included calculating norms and errors for correlation and transition matrices. One study assessed the statistical significance of correlation coefficients. Evaluations of categorical variables, particularly their comparison with continuous variables, were less frequent.

Multivariate resemblance evaluation (Figure 4, panel 3) involved assessing the similarity of distributions across multiple variables between the synthetic and original data. Qualitative paradigms, employed in 12 studies (36%), primarily utilized visualizations of dimensionality reduction techniques, such as principal component analysis and t-distributed stochastic neighbor embedding. Quantitative paradigms, implemented in 11 studies (33%), included metrics such as clinician-assigned scores and results of model-based evaluations, e.g., area under the curve (AUC) and classification accuracy. These models, used in five studies (15%), included random forests, LSTM networks, and "Other" models, such as naïve classifiers, transfer learning, and fine-tuning-based discriminators. Additionally, three studies (9%) employed statistical tests, including the KS-test, maximum mean discrepancy across multiple variables, and t-test to infer discriminator performance.

Temporal resemblance assessment (Figure 4, panel 4) between synthetic and original variables involved evaluating distributional behavior over time. This was qualitatively assessed in 21 studies (64%), primarily through visualizing mean values over time using line plots or consecutive measurements with scatter plots. "Other" qualitative approaches involved visualizing underlying network structures and using tile plots. Quantitative paradigms were used in 13 studies (39%), including the calculation of means, standard deviations for visit intervals, and autocorrelation functions. "Other" quantitative approaches included calculating the RMSE of autocorrelation functions, a singular value-based latent temporal statistic, bigram sequences, and metrics assessing the loss of temporal information, trends, and cycles. One study applied a model-based approach, predicting the next value based on previous ones using an RNN.

### *3.5.2 Utility*

Utility was evaluated in 26 studies (72%), focusing on either statistical inference or prediction performance, including classification tasks. These domains were divided into the same four paradigms as for resemblance. Figure 5 provides an overview of the utility assessment approaches.

Eight of these studies (31%) evaluated the similarity of statistical inferences between the synthetic and original data (Figure 5, panel 1). Seven of them employed both qualitative and model-based approaches, while one focused on comparing the results of the Mann-Whitney U test. Qualitative paradigms were primarily used for visualizing model parameter estimates and confidence interval overlap. Additionally, "Other" quantitative approaches included evaluating ratios of standard errors and effect size. The models utilized included linear, logistic, and mixed-effects regression, factor analysis, general estimation equations, and causal effect models.

Eighteen studies (69%) assessed predictive performance (Figure 5, panel 2), all using model-based paradigms. These studies trained models with both synthetic and original data, comparing performance on a test set. The models employed included logistic regression, random forest, LSTM, and other methods such as support vector machines, RNN, and batch-constrained Q-learning. Ten studies used qualitative approaches to visualize model performance. Model performance metrics—such as accuracy, AUC, recall, and $F_1$-score—were classified under "Other" quantitative approaches and compared in 13 studies. Two studies used t-tests to statistically assess differences in prediction performance between models.

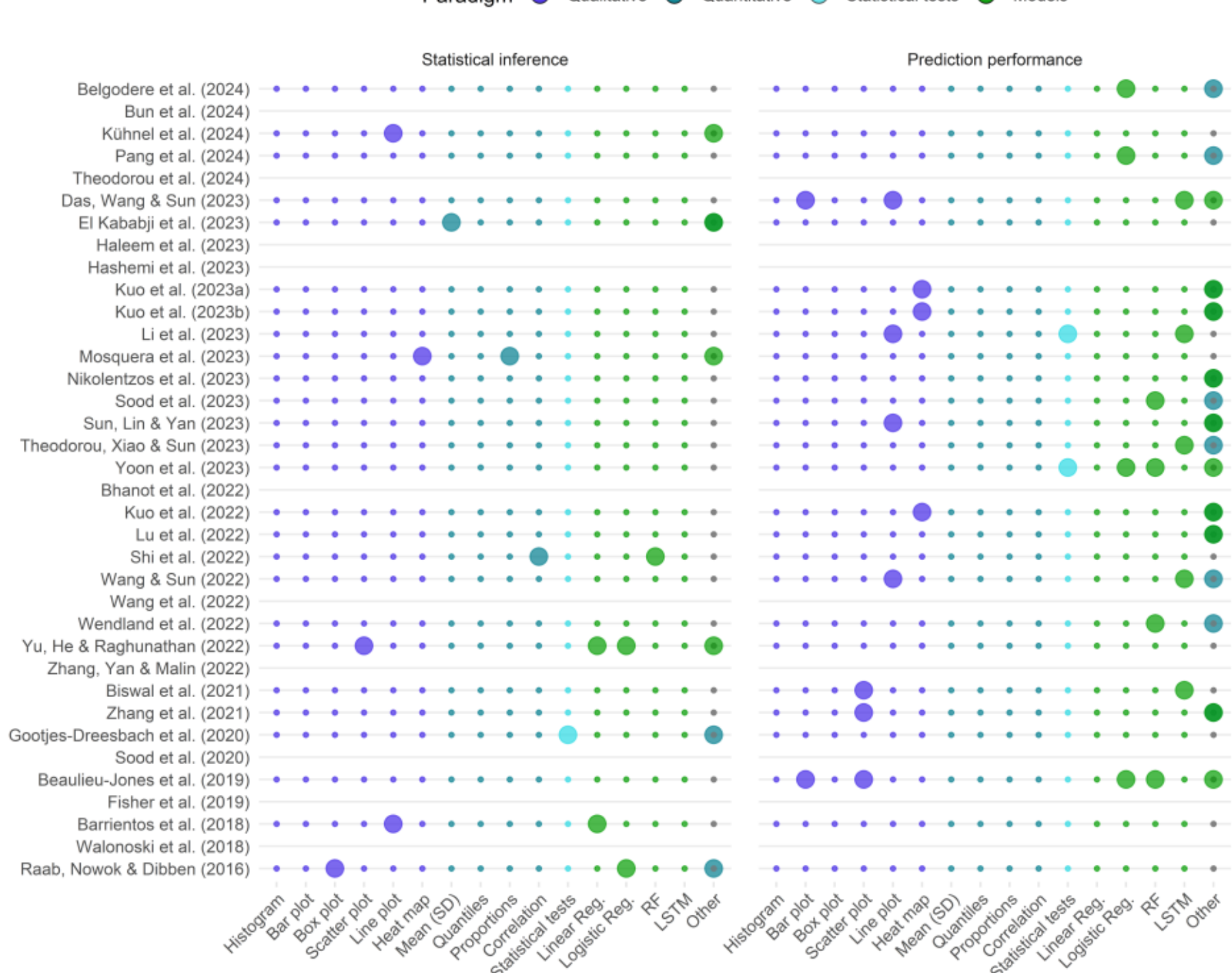

Figure 5: Approaches used to evaluate synthetic data utility in comparison to the original data. The figure illustrates the application of various approaches (shown on the x-axis) for assessing the utility between the synthetic and original data in the included studies, with the larger circles indicating the approach was used. The evaluation of utility encompasses two domains: similarity of statistical inference and prediction performance, including classification. Furthermore, the approaches used for this assessment were categorized into qualitative, quantitative, model-based, and statistical test-based paradigms. It should be noted that a study may encompass multiple assessments that belong to the same approach. The approach "Histogram" also includes density plots, and the approach "Box plot" includes violin plots. Statistical tests and approaches labeled as "Other" are discussed in the main text. SD: standard deviation; Reg.: regression; RF: random forests; LSTM: long short-term memory; NA: not applicable, did not compare synthetic and original data.

### *3.5.3 Privacy*

Privacy concerns, distinct from the privacy preservation techniques employed within the SDG method itself, were addressed in 19 studies (53%), as outlined in Table 4. These studies utilized a range of threat models to address various privacy concerns, with six studies examining two separate aspects.

Table 4: Approaches for evaluating privacy in synthetic data. The table summarizes studies that assessed various disclosure risks. Attribute disclosure occurs when sensitive or private information about an individual can be inferred from synthetic data. Identity disclosure happens when an individual's identity is linked to their data. Membership disclosure involves the identification of an individual's participation in the original dataset.

| Study | Attribute disclosure | Identity disclosure | Membership disclosure |
|---|---|---|---|
| Belgodere et al. [4] | | | ✓ |
| Pang et al. [52] | ✓ | | ✓ |
| Das, Wang & Sun [12] | ✓ | | ✓ |
| El Kababji et al. [34] | ✓ | | ✓ |
| Hashemi et al. [30] | | ✓ | |
| Kuo et al. [36] | | ✓ | |
| Kuo et al. [37] | | ✓ | |
| Li et al. [41] | | | ✓ |
| Mosquera et al. [45] | ✓ | | |
| Nikolentzos et al. [48] | | | ✓ |
| Sun, Lin & Yan [66] | ✓ | | ✓ |
| Theodorou, Xiao & Sun [70] | ✓ | | ✓ |
| Yoon et al. [79] | | | ✓ |
| Kuo et al.[38] | | ✓ | |
| Shi et al. [59] | | ✓ | |
| Wang & Sun [75] | | | ✓ |

| Study | Attribute disclosure | Identity disclosure | Membership disclosure |
|---|---|---|---|
| Biswal et al. [6] | | | ✓ |
| Zhang et al. [81] | ✓ | | ✓ |
| Barrientos et al. [2] | ✓ | | |

The majority of these studies (63%) focused on membership disclosure, defined as the inadvertent or unauthorized identification of an individual's inclusion in the original (training) dataset. Common approaches included distance-based attacks, where the similarity between synthetic and training data was compared to that between synthetic data and a hold-out test set using predefined metrics, or classifier-based attacks, where models were trained to differentiate between real and synthetic samples. Synthetic data were deemed "safe" if the distances were comparable or if the classifier failed to distinguish synthetic from test observations within the applied accuracy threshold.

Attribute disclosure, the unintended or unauthorized exposure of specific individual attributes (variables), was the second most frequently assessed aspect of privacy, examined in eight studies (42%). The threat models involved quantifying attribute matches between original and synthetic datasets using a defined distance metric, as well as training models on both synthetic and original data to assess the likelihood of observing particular attribute combinations or values. Attribute disclosure was considered to occur when the predicted probability exceeded a predefined threshold.

Identity disclosure, defined as the unintended revelation of an individual's identity, was assessed in five (26%) studies. Each study measured the Euclidean distance between synthetic and original data, considering re-identification to occur when the distance was less than a predefined threshold. However, the method of calculating this distance varied: some studies weighted the distances and compared them to a dataset with one individual excluded, similar to leave-one-out cross-validation, while others calculated distances directly between the datasets.

# 4 DISCUSSION

We identified 36 studies presenting methods able to generate synthetic longitudinal patient data, with the majority published in 2023. Of the 39 primary methods, 11 were mentioned in earlier medical SDG reviews [19, 20, 32, 43, 45, 47], with the number in parenthesis indicating how many of the six publications included each method: AC-GAN (5), ADS-GAN (1), CRBM (2), EHR-M-GAN (1), EVA (1), HealthGAN (5), LS-EHR (1), SynTEG (3), Synthea (2), synthpop (1), and VAMBN (1). The difference in the number of methods identified in our review compared to earlier reviews cannot be fully explained by our inclusion of more recent publications as a significant portion of the methods from overlapping years were either not recognized or were mentioned in only one of the earlier reviews. This likely results from variations in the databases and search strategies used, as well as the fact that some earlier reviews focused on specific methods like GANs or, except for [45], on different data types.

In addition to identifying a larger number of methods, the current review differs from the earlier ones by providing a comprehensive and in-depth discussion and classification of the identified SDG methods from the perspective of longitudinal data. We specifically investigate how these methods address the temporal dimension of LD, their ability to generate unbalanced LD, and the approaches used to evaluate temporal preservation—topics that have not been properly addressed in previous reviews. Longitudinal data, in general, remain underexplored in the SDG literature, where terms like "time-series", "trajectory", or simply "EHRs" are often used interchangeably, highlighting the challenge of distinguishing between various time-dependent data types. Our findings reflect this difficulty, with approximately two-thirds of the methods focusing on generating standard longitudinal data and the rest on sequential or trajectory-based data (see Section 2.7). Given the overlap of these latter two types with time series, readers seeking methods for generating such data may find it useful to consult also the literature on synthetic time series generation.

Most identified methods were deep learning (DL) models. These models capture complex patterns from data without requiring strict distributional assumptions. However, DL models typically entail numerous training parameters and demand substantial sample sizes, limiting their use with small datasets. Dealing with multiple variable types is challenging and often necessitates variable encoding and normalization, reducing information and increasing dimensionality. Moreover, the effectiveness of DL models in generating longitudinal data relies heavily on their ability to discern patterns within the input data, yet they typically struggle with missingness and generating unbalanced observations. In the context of synthetic LD generation, a step forward would involve developing and integrating components specifically designed to preserve temporal structures and generate unbalanced data, as seen in two novel approaches, cLSTM [45] and EHR-Safe [79], identified in this review.

Language models have made progress in addressing missingness and unbalance in LD. However, it remains unclear whether missing values arise from the inherent unbalanced structure of LD or if models are learning patterns of missingness and its distribution. Additionally, our empirical tests on some of the identified methods reveal that these models can generate values absent in the training data due to the vocabulary being constructed from the entire dataset. Consequently, these generated values do not reflect the proportions observed in the population, highlighting the importance of selecting representative training data for LM-based SDG. Furthermore, the current implementations often require sample sizes in the tens of thousands and considerable computational resources, far exceeding those of traditional DL models. This makes them more suitable for large hospital databases than for study-specific datasets. Additionally, many implementations lack privacy-preserving mechanisms, which further limits their applicability.

While nearly all studies assessed resemblance, only every fifth evaluated all four dimensions: univariate, bivariate, multivariate, and temporal preservation. It is encouraging to see that the evaluation of temporal preservation, which we consider essential when generating synthetic LD, has become more prevalent in recent studies (Figure 4, panel 4). However, the majority of studies still relied on subjective qualitative techniques. Temporal preservation could be more rigorously assessed using quantitative methods, such as comparing autocorrelations or transition matrices. Moreover, the relationship between categorical and numerical variables remains largely unexplored. The evaluation of utility was even less frequent, with only three-fourths of studies addressing it, and there was a clear distinction between statistical inference and prediction tasks, with most studies focusing exclusively on the latter and none addressing both. This raises concerns about the reliability of statistical utility of the synthetic data generated by these methods.

We observed no clear trends in the integration of privacy-preserving techniques within SDG, as only one in five methods implemented such measures. Furthermore, only half of the studies included a privacy evaluation, a pattern also noted by [46]. This limited focus on privacy may stem from the assumption that generating random data inherently ensures privacy, the challenges associated with implementing and evaluating privacy protections effectively, and the lack of established guidelines for defining and measuring privacy thresholds. The included studies often utilized similar approaches to assess different disclosure scenarios, making it difficult to ascertain which aspects of privacy were actually evaluated. The use of varied terminologies for the same concepts added further complexity. Nevertheless, it is crucial to assess SD for potential disclosure risks, particularly identity disclosure, as it often leads to subsequent disclosures. This holds true even when employing privacy-preserving methods like DP, as certain DP implementations may not consistently adhere to theoretical privacy bounds [64]. Moreover, the practical interpretation of the privacy budget parameter is not straightforward [40] and choosing the correct privacy budget is challenging. For large census data like EHRs of hospital districts, the value of assessing membership disclosure is unclear since they typically encompass the entire regional population, making it reasonable to assume that everyone in the area is likely included in the original data. Nevertheless, an inherent tradeoff persists between privacy and utility, necessitating finding a balance between these two factors.

Of particular concern is that only one in ten studies comprehensively evaluated resemblance, at least one dimension of utility, and one aspect of privacy, casting doubt on the real-world applicability of the identified methods. This gap is likely due to the absence of a standardized evaluation framework, as highlighted by previous reviews [19, 20, 32, 39, 43, 47]. While initial attempts to develop such frameworks have been made [31, 78], they are largely based on cross-sectional data and do not adequately address other data types, such as LD. Our classification of resemblance, utility and privacy provides a foundation for developing a more comprehensive evaluation framework. Although some approaches are applicable to many different data types, our categorization facilitates the integration of data-specific (e.g., temporal structure) and task-specific utility metrics, accommodating both statistical and machine learning perspectives. However, the detailed specification of these metrics remains a topic for future research. Finally, a crucial part of any evaluation is to use independent replications and average the metrics across them. If only a single synthetic dataset is used, any observed outcome is possible to have occurred by chance alone. For reliable conclusions, the required number of replicates is likely in tens/hundreds. Surprisingly, only one in ten studies employed more than one replication.

Lastly, none of the previous reviews [19, 20, 32, 39, 43, 45, 47] systematically assessed risk of bias or reporting quality, although [32] recognized the potential for biases. Thus, our review appears to be the first in this regard. Despite our structured framework and validation by all authors, these assessments remain somewhat subjective, a recognized characteristic even within established frameworks [61]. Approximately every third publication lacked a transparent description of comparison procedures (risk of performance bias), and selective reporting (risk of reporting bias) was present almost in half of the studies. Furthermore, like [20], we identified inadequacies in the training process descriptions and source code availability. Our findings are not meant to criticize any single study, and any shortcomings are most likely oversights. The review's findings should be regarded as overarching recommendations for improving the transparency of SDG research.

### 4.1 Limitations

Although our review, to our best knowledge, is the first systematic literature review on SDG that adheres thoroughly to the PRISMA guidelines (checklist in Appendix A.8), including the preparation and submission of a review protocol (CRD42021259232) to the international systematic review registry PROSPERO [85], it is important to acknowledge its inherent constraints.

First, due to the lack of unambiguous definitions of synthetic and longitudinal data and the diverse evolution of SDG across various fields, formulating a definitive search algorithm was challenging, leading to several false positives during the screening phase (see Figure 2). Despite our efforts to encompass synonyms, omitting relevant publications remains possible. Nevertheless, given the large number of screened publications and exhaustive citation searching, coupled with our accurate identification of intersecting publications observed in the prior reviews, we are confident in the comprehensiveness of the literature included.

Second, longitudinal data analysis has mainly been used in medical statistics, while SDG methods derive predominantly from computer science and associated applications. This discrepancy poses challenges in assessing the applicability of SDG methods for LD generation. For instance, the notion of unbalanced data, though well-established in statistics, has received limited attention in computer science, resulting in its underrepresentation in SDG research. This likely explains why the respective information was not available in the included studies (Table 3).

Third, it is crucial to note that scientific advancements continue, and this review is confined to data accessible until May 2024. Since most of the methods identified in this review were published from 2022 onward, it is likely that new methods have already been introduced. Our findings suggest that these newer methods are likely addressing issues related to the unbalanced nature and missingness of LD, and we anticipate that privacy-preserving mechanisms are being integrated into these approaches. Nevertheless, we believe that our work provides a solid foundation by presenting 39 methods—this number increases to 66 when including the 27 reference methods—offering researchers and data controllers a comprehensive selection of approaches to meet their specific needs, while also highlighting current shortcomings and suggesting practical areas for improvement in future method development.

### 4.2 Conclusion

Our review identified 39 methods for generating synthetic longitudinal patient data (RQ1), addressing various challenges associated with longitudinal data (RQ2). Only EHR-Safe [79], DSG [26], HALO [70] and cLSTM [45] addressed all challenges simultaneously, but none incorporated privacy-preserving mechanisms, and all relied on resource-intensive deep learning or language models. Furthermore, the quality of evaluating synthetic data varied significantly across studies (RQ3), leaving the methods' real-world applicability uncertain, especially across diverse datasets. These findings highlight a critical gap and underscore the need for developing more robust, efficient, and privacy-preserving synthetic data generation methods for longitudinal data.

Due to the many different forms of LD, no single method is likely to cover all cases and more focused approaches are advised. For example, for standard longitudinal data one could consider a hybrid approach where static covariates are generated using an assumption-free generative model and unbalanced time-varying responses are modelled conditionally on the covariates using hierarchical models that acknowledge the missingness and time-dependence in the measurements.

The observed heterogeneity in the evaluation approaches across studies presents a significant challenge in conducting comparisons between methods and their applicability in practice. While creating standardized evaluation criteria would enhance method assessment, recognizing the importance of tailored approaches for various applications is crucial. Establishing a standardized evaluation framework offers a chance for interdisciplinary collaboration among medicine, statistics, and computer science. Our categorization of different evaluation approaches for assessing resemblance, utility and privacy provides a robust basis for subsequent research.

Lastly, further research is needed to address privacy concerns surrounding SD, along with clear guidance from data protection authorities, such as official standards for defining privacy thresholds for data publication. Currently, there is a gap between the development of SDG methods and the availability of synthetic patient data and platforms to facilitate their use. Bridging this gap will require collaboration among method developers, medical practitioners and policymakers as the directives will require empirical support and new methods should be developed with practical feasibility in mind.

## ACKNOWLEDGMENTS

We thank Antti Airola, Martin Closter Jespersen, Henning Langberg and Arho Virkki for their roles in the systematic review team.

This work was supported by the Novo Nordisk Foundation (grant number NNF19SA0059129), the Finnish Cultural Foundation (grant number 00220801) and the Research Council of Finland (grant numbers 335077, 347501 and 353769). The funding agencies were not involved in study design; the collection, analysis and interpretation of data; the writing of the report; nor in the decision to submit the article for publication.

The review protocol was registered with PROSPERO (registration number CRD42021259232) on July 5, 2021. Subsequently, the protocol was amended twice, on January 25, 2022, and March 9, 2023, respectively. The rationale for each modification is detailed in the corresponding section of the amended protocol, which is available at https://www.crd.york.ac.uk/prospero/display_record.php?ID=CRD42021259232.

The data used to derive the results and conclusions of this systematic review are available upon request.

# A APPENDICES

## A.1 Search algorithms

### *A.1.1 Web of Science (Core Collection)*

Search date 2021-06-11, 3795 hits

```
#1 TS = ((synthetic OR artificial)
NEAR/3 (*data* OR record*))
AND TS = ((generat* OR produc* OR simula*))
AND TS = ((longitudinal OR correl* OR panel
           OR repeat* OR follow-up
           OR multivariate OR lifespan*
           OR traject* OR health*
           OR medical OR patient))
NOT TS = (aperture OR insemination OR seism*)
```

```
AND LA = (English)
AND DT = (Article OR Abstract of Published Item
          OR Book OR Book Chapter OR Data Paper
          OR Early Access OR Proceedings Paper
          OR Review OR Software Review)
```

Search date 2022-11-22, 1734 hits

```
#2 TS = ((synthetic OR artificial)
NEAR/3 (*data* OR record*))
AND TS = ((generat* OR produc* OR simula*))
AND TS = ((longitudinal OR correl* OR panel
           OR repeat* OR follow-up
           OR multivariate OR lifespan*
           OR traject* OR health*
           OR medical OR patient))
NOT TS = (aperture OR insemination OR seism*)
AND LA = (English)
AND DT = (Article OR Abstract of Published Item
          OR Book OR Book Chapter OR Data Paper
          OR Early Access OR Proceedings Paper
          OR Review OR Software Review)
NOT #1
```

Search date 2024-05-21, 1278 hits
Like 2022-11-22 but limited to results published after 2022-11-22

*A.1.2 Embase (1947 onwards)*

Search date 2021-06-11, 504 hits

```
#1 (((synthetic OR artificial)
NEAR/3 (data OR record*)):ti,ab,kw)
AND (generat* OR produc* OR simula*):ti,ab,kw
AND (longitudinal OR correl* OR panel
OR repeat* OR ’follow?up’
OR multivariate OR lifespan*
OR traject* OR health* OR medical
```

```
OR patient):ti,ab,kw
AND ([article]/lim OR [article in press]/lim
OR [conference paper]/lim
OR [conference review]/lim
OR [data papers]/lim OR [letter]/lim
OR [note]/lim OR [review]/lim
OR [short survey]/lim)
AND [english]/lim
AND [embase]/lim
```

Search date 2022-11-22, 326 hits

```
(((synthetic OR artificial)
NEAR/3 (data* OR record* OR microdata*)):ti,ab,kw)
AND (generat* OR produc* OR simula*):ti,ab,kw
AND (longitudinal OR correl* OR panel OR repeat*
               OR 'follow?up' OR multivariate OR lifespan*
               OR traject* OR health* OR medical OR patient):ti,ab,kw
NOT (aperture OR insemination OR seism*):ti,ab,kw
AND ([article]/lim OR [article in press]/lim
               OR [conference paper]/lim OR [conference review]/lim
               OR [data papers]/lim OR [letter]/lim OR [note]/lim
               OR [review]/lim OR [short survey]/lim)
AND [english]/lim NOT #1
```

Search date 2024-05-21, 314 hits
Like 2022-11-22 but limited to results published after 2022-11-22

*A.1.3 MEDLINE (Ovid interface, 1946 onwards)*
Search date 2021-06-12, 574 hits

```
#1 (((synthetic or artificial)
    adj3 (data or record*))
    and (generat* or produc* or simula*)
    and (longitudinal or correl* or panel
        or repeat* or 'follow up'
        or multivariate or lifespan*
```

```
        or traject* or health* or medical
        or patient)).ti,ab,kf.
 #2     limit #1 to ((english language or english)
    and (classical article or clinical conference
        or comparative study or congress
        or english abstract or evaluation study
        or festschrift or government publication
        or historical article
        or introductory journal article
        or journal article
        or letter or preprint or "review"
        or "systematic review" or technical report
        or validation study))
```

Search date 2022-11-22, 402 hits (contains duplicates with the previous search because the time range could not be specified more precisely)

```
#3 (((synthetic or artificial)
    adj3 (data* or record* or microdata*))
    and (generat* or produc* or simula*)
    and (longitudinal or correl* or panel
        or repeat* or 'follow up'
        or multivariate or lifespan*
        or traject* or health* or medical
        or patient)
    not (aperture OR insemination OR seism*)).ti,ab,kf
#4 limit #3 to ((english language or english)
    and (classical article or clinical conference
    or comparative study or congress
    or english abstract or evaluation study
    or festschrift or government publication
    or historical article
    or introductory journal article
    or journal article or letter or preprint
    or "review" or "systematic review"
```

```
    or technical report or validation study))
#5 limit #3 not #2
```

Search date 2024-05-21, 355 hits
Like 2022-11-22 search.

*A.1.4 Google Scholar (Publis or Perish software, 1000 first hits)*

Search date 2021-06-18, 980 hits

```
("synthetic data" OR "artificial data")
AND (generat* OR priduc* OR simula*)
AND (longitudinal OR correl* OR panel
    OR repeat* OR "follow up" OR "follow-up"
    OR "multivariate OR lifespan* OR traject*
    OR health* OR medical OR patient)
```

The search was not repeated because Google Scholar flagged the software as a bot, leading to a temporary ban.

*A.1.5 arXiv*

Open-source metadata were downloaded from Kaggle [18] and R software (version 4.2.2) [71] was used to extract the relevant articles. The source code is presented below.

Search date 2022-11-22, 628 hits

```
# libraries
library(jsonlite)
library(data.table)
library(synthesisr)

# importing ArXiv results
arxiv <- stream_in(file(paste0(getwd(), "/articles/source_searches/arxiv-metadata-oai-snapshot.json")))
arxiv <- as.data.table(arxiv)

# regex developed according to database search queries
# synthetic data
search_data <-"\\b(synthetic|artificial)(?:\\W+\\w+){0,3}?\\W?(\\S*data|record\\S*)\\b"
```

```
# inclusion criteria
search_gener <- "(generat|produc|simula)"
search_type                  <-                  "(longitudinal|correl|panel|repeat|follow-
up|multivariate|lifespan|traject|health|medical|patient)"

# exclusion criteria
search_excl <- "(aperture|insemination|seism)"

# grepping abstracts according to criteria
arxiv_results_1 <- arxiv[grepl(search_data, abstract, ignore.case = T, perl = T)]
arxiv_results_2 <- arxiv_results_1[grepl(search_gener, abstract, ignore.case = T, perl = T)]
arxiv_results_3 <- arxiv_results_2[grepl(search_type, abstract, ignore.case = T, perl = T)]
arxiv_results_4 <- arxiv_results_3[!grepl(search_excl, abstract, ignore.case = T, perl = T)]

# modifying data for export
arxiv_results_4[, source_type := ifelse(is.na(`journal-ref`), "UNPB", "JOUR")]
arxiv_results_4[is.na(`journal-ref`), `journal-ref` := paste0("arXiv preprint arXiv:", id)]
arxiv_results_4[, year := year(update_date)]

setnames(arxiv_results_4, "journal-ref" , "journal")
setnames(arxiv_results_4, "update_date" , "date_generated")
setnames(arxiv_results_4, "authors" , "author")

arxiv_results_4[, c("id", "submitter", "comments", "report-no",
                    "categories", "license", "versions", "authors_parsed") := NULL]

setcolorder(arxiv_results_4, c("date_generated", "source_type", "author",
                               "year", "title", "journal", "doi"))

arxiv_results_4[, author := gsub(",", " and", author)]
arxiv_results_4[, author := gsub("\n", "", author)]
arxiv_results_4[, author := gsub("\\\\", "", author)]
arxiv_results_4[, author := gsub("\\\"", "", author)]
arxiv_results_4[, author := gsub("[(]\\d[)](\\W?and)?", "and", author)]
```

```
# exporting as ris file
write_refs(as.data.frame(arxiv_results_4),   format   =   "ris",   file   =   paste0(getwd(),
"/articles/source_searches/arxiv_results.ris"))
```

Search date 2024-05-21, 417 hits
Similar to 2022-11-22, limited to results published after 2022-11-22.

### A.2 Selection process

*A.2.1 Abstract screening chart*

Each included search result was screened by KP and JV independently using Rayyan [67] and the flowchart presented below. The process started at the top of the chart (Start) and progressed in the directions indicated by the arrows, depending on the selection. The terminations of the process and the selection of the search result (include, maybe, exclude) are indicated in bold.

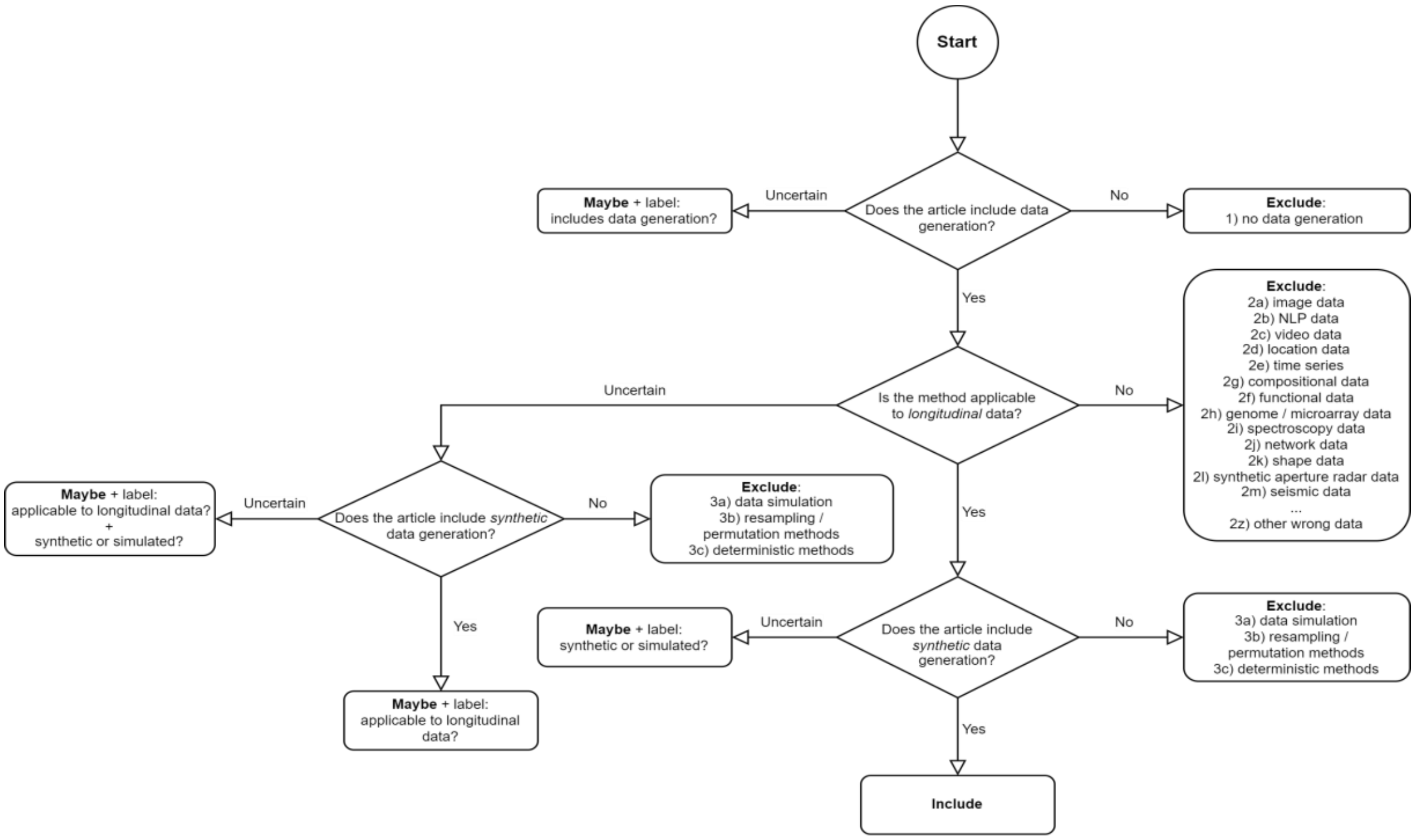

*A.2.2 Full-text screening chart*

Full texts of each publication that was included after screening the titles and abstracts was screened by KP and JV independently using Rayyan [67] and the flowchart presented below. The process started at the top of the chart (Start) and progressed in the directions indicated by the arrows, depending on the selection. The actions and terminations of the process and the selection of the search result (include, maybe, exclude) are indicated in bold

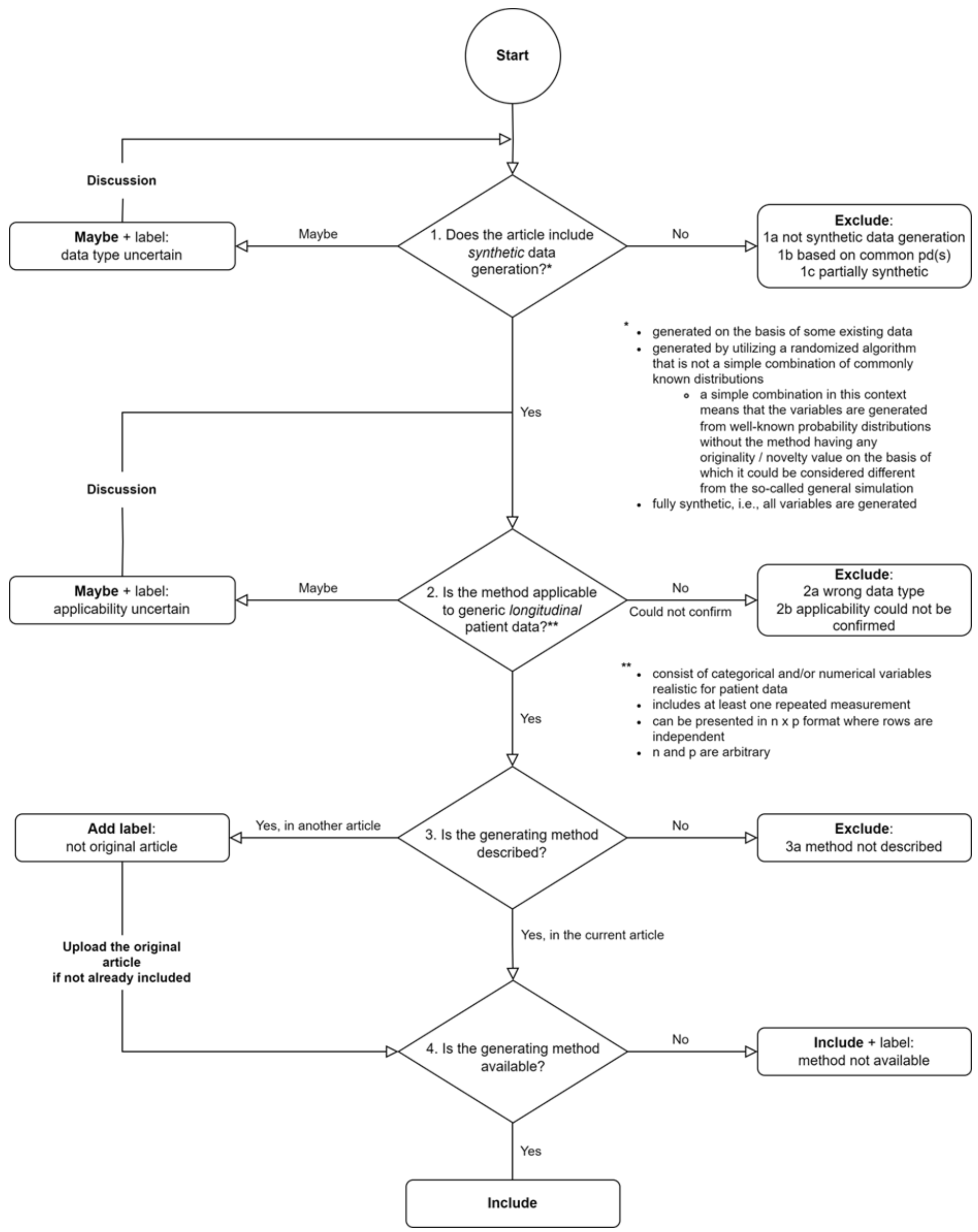

### A.3 Data collection process

Data were collected and managed by the corresponding author using a structured form designed in REDCap electronic data capturing tools hosted at University of Turku [35, 36]. The forms are presented below.

*A.3.1 Literature information*

# Literature information

*Page 1*

Please complete the survey below.

| | |
|---|---|
| Type of publication | ◯ Journal article<br>◯ Poster<br>◯ Conference paper<br>◯ Book chapter<br>◯ Dissertation or thesis<br>◯ Research report<br>◯ Review article<br>◯ Other |
| Name the other publication type | ______________________________ |
| Authors | ______________________________________ |
| Year | ______________________________ |
| Title | ______________________________ |
| Publication platform (journal, conference, book...) | ______________________________<br>(Give the name of the journal/conference etc.) |
| Volume | ______________________________ |
| Issue | ______________________________ |
| Page numbers | ______________________________ |
| Is the publication peer-reviewed? | ◯ Yes<br>◯ No |
| What was the purpose of the study? | ______________________________________<br>(In this context, study refers to the article) |

11.08.2023 11:27 projectredcap.org REDCap®

*A.3.2 Method characteristics*

# Method characteristics

Please complete the survey below.

| **Basic information** | |
|---|---|
| Type of the method | ☐ Generative adversarial network<br>☐ Recurrent neural network<br>☐ Auto-encoder (variational or other)<br>☐ Bayesian network<br>☐ Hidden Markov model<br>☐ Density estimation<br>☐ Imputation method<br>☐ Dimensionality reduction<br>☐ Data partitioning<br>☐ Decision tree (classification, regression)<br>☐ Posterior predictive sampling<br>☐ Clustering<br>☐ Other deep learning method<br>☐ Other |
| Type of the method (other/other deep learning) | ______________________________ |
| Is expert knowledge required/used in the method? | ◯ Yes<br>◯ No |
| How is expert knowledge needed? | ____________________________________ |
| Describe the method as concisely as possible | ____________________________________ |
| Programming language | ☐ R<br>☐ Python<br>☐ C++<br>☐ Java<br>☐ Scala<br>☐ Julia<br>☐ Fortran(77/90/95/...)<br>☐ Matlab/Octave<br>☐ SAS<br>☐ Other<br>☐ Not specified |
| Programming language (other) | ______________________________ |
| Is the pseudocode of the method presented? | ◯ Yes<br>◯ No |
| Is the source code/software provided? | ◯ Yes<br>◯ No<br>◯ Upon request |

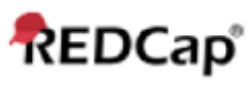

| | |
|---|---|
| Method's source code location | ______________________________ (e.g. URL) |
| Is the software | ◯ Library<br>◯ Standalone software<br>◯ Other<br>◯ Not specified |
| Define the other software type | ______________________________ |
| Is the software used to apply the method free? | ◯ Yes<br>◯ No<br>◯ Not specified |
| Software licence | ______________________________ |
| Was the method | ☐ Originally desinged for longitudinal data<br>☐ Altered/modified for longitudinal data<br>☐ Implemented to longitudinal data without any modifications |
| How the method model/approaches longitudinal data | ______________________________ |

**Used system and complexity (requirements)**

| | |
|---|---|
| Does the article mention anything about the used system or its requirements? | ◯ Yes<br>◯ No |
| Operating system | ______________________________ |
| Other system requirements or used system information | ______________________________ |
| Method's running time in terms of the input size (Big O notation) if reported | ______________________________ |

**Input (original) and output (synthetic) data properties**

| | |
|---|---|
| Is the method capable of handling unbalanced longitudinal data? | ◯ Yes (number of time points / timing / spacing for intervals is different for different subjects)<br>◯ Yes (some variables are collected less often than others, but still for everyone at the same time point)<br>◯ No<br>◯ Not specified |
| Is the method capable of generating unbalanced longitudinal data? | ◯ Yes (number of time points / timing / spacing for intervals is different for different subjects)<br>◯ Yes (some variables are collected less often than others, but still for everyone at the same time point)<br>◯ No<br>◯ Not specified |
| The method is capable of | ☐ Handling categorical original data<br>☐ Handling numerical original data<br>☐ Generating categorical synthetic data<br>☐ Generating numerical synthetic data<br>(numerical = continuous / interval, categorical = binary / multiclass) |
| The numerical data values generated | ☐ Will not necessarily fall within the corresponding range in the original data set<br>☐ Will fall within the corresponding range in the original data set<br>☐ Will be replicates of values in the original data set<br>☐ Not specified |
| Is the method capable of handling missing values in the original data? | ◯ Yes<br>◯ No<br>◯ Not specified |
| Is the method capable of producing missing values for synthetic data? | ◯ Yes<br>◯ No<br>◯ Not specified |

*A.3.3 Method evaluation*

# Method performance evaluation

Please complete the survey below.

| **Data used to generate synthetic data, i.e., original or input data** | |
|---|---|
| Synthetic data was generated based on | ☐ Real-world data<br>☐ Simulated data<br>☐ Synthetic data<br>☐ Other<br>(i.e., what type/form was the original data?) |
| Give the name of the data set(s) | ____________________________________<br>(Separate the names with a comma) |
| What kind of data was used? Separate different data sets with a comma. | ____________________________________<br>(e.g., patient data, other data related to people, non-human data) |
| Is the used data set(s) available? | ☐ Publicly<br>☐ Upon request<br>☐ No |
| Give the source(s) of the available data set(s) | ____________________________________<br>(If multiple, in same order as given the data sets above) |
| The number of independent observations (subjects) in the input data set(s). If multiple data sets, separate with a comma. | ________________________________<br>(If not reported or recoverable, write 'Not specified') |
| The number of variables in the input data set, including variables with repeated measurements. If multiple data sets, separate with a comma. | ________________________________<br>(If not reported or recoverable, write 'Not specified') |
| Number of variables with repeated measurements (subset of the total number of variables). | ________________________________ |
| Number of repeated measurements. For unbalanced data or varying ranges, give the range [min, max]. If multiple data sets are used, separate them with a comma. | ________________________________ |
| Number of categorical variables. If multiple data sets are used, separate them with a comma. | ________________________________<br>(Options: 0,1,,2,..., not specified) |
| Number of numerical variables. If multiple data sets are used, separate them with a comma. | ________________________________<br>(Options: 0,1,,2,..., not specified) |

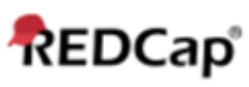

Is the pattern of missingness similar to the original data?

◯ Yes
◯ No
◯ Not specified

---

Does the method have other limitations or requirements for the original data that have not already been mentioned?

◯ Yes
◯ No
(e.g. input data have to be scaled)

---

Describe the requirements / limitations regarding to original data

__________________________________

## Evaluation setup of the generated synthetic data

Was the variable(s) in the synthetic data with repeated measurement treated as

◯ Response
◯ Explanatory
◯ Both
◯ Not specified
(Option "Not specified" should be used only if the article is otherwise relevant and the nature of the variable cannot be determined even through discussion.)

---

The evaluation of the generated synthetic data was based on

☐ Qualitative assessment
☐ Quantitative assessment
☐ Other
(Select all suitable options)

---

Describe the other approach used to evaluate the synthetic data and/or the method

__________________________________

---

The evaluation of the generated synthetic data was based on

☐ A single repetition (i.e., the assessment is based on a single generated synthetic data set)
☐ A small amount of repetitions (i.e., multiple data sets)(< 50)
☐ A large amount of repetitions (>= 50)
(Select all suitable options)

---

Was any of the following used to describe or evaluate the generated synthetic data and/or the method

☐ Descriptive statistics
☐ Statistical inference
☐ Prediction/classification (synthetic vs. real)
☐ Prediction/classification (some other variable)
☐ Privacy
☐ Externally assessed realism
(Select all suitable options)

---

Was the generated synthetic data evaluated

☐ Against (resamples) original data
☐ Against other simulated data
☐ Against other real-world data (public / private)
☐ Against another synthetic data set(s) generated by the same method (e.g., using different parameters)
☐ Against another synthetic data set(s) generated by a different method or methods
☐ No comparisons to other data or methods were made (i.e., a single data set was generated)
☐ Other
(Select all suitable options)

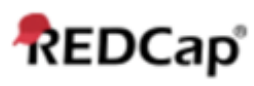

Name the other methods used in the comparison

__________________________________

Describe the other approach to used to evaluate the generated synthetic data (in terms of data)

__________________________________

Describe the simulation approach.

__________________________________

Was the training process(es) described and/or available in the source code?

○ Yes
○ Partially
○ No

What was lacking from the training process description if it was only partially described?

__________________________________

## Qualitative methods used to characterize and/or evaluate the generated synthetic data set(s)

Specify all qualitative methods (e.g., figures) used to describe and/or evaluate the generated synthetic data set(s)

__________________________________

## Descriptive methods used to characterize and/or evaluate the generated synthetic data set(s)

Specify all descriptive statistics (e.g., measures, estimates) used to describe and/or evaluate the generated synthetic data set(s)

__________________________________

## Inferential statistics used to evaluate the generated synthetic data set(s)

Specify all inferential statistics (e.g., tests, models) used to evaluate the generated synthetic data set(s)

__________________________________

## Predictive and classification approaches used to evaluate the generated synthetic data set(s)

Specify all the predictive and classification approaches (e.g., models, accuracy measures) used to evaluate the generated synthetic data set(s) in terms of synthetic data performance

__________________________________

## Privacy of the method and the generated synthetic data set(s)

Was differential privacy used to enhance/secure the privacy of the generated synthetic data?

○ Yes
○ No
(e.g., as a part of the method / applied post-hoc)

What was the epsilon used? If multiple epsilons were used, separate them with a comma

____________________________

Specify delta if applicable. If value not specified, write not specified.

____________________________

| | |
|---|---|
| Was any of the following used to test the privacy of the synthetic data? | ☐ Membership attack / identity disclosure<br>☐ Attribute disclosure<br>☐ Inferential disclosure<br>☐ Other<br>☐ No other approaches were used |
| Specify how the privacy of the method and/or the generated synthetic data set(s) was addressed: specify the approach (e.g., distinguishing records with a model) and parameters used (other than DP asked previously) if reported or write a summary of the authors' discussion on the subject if no specific approach was used. | __________________________ |

**Externally assessed realism**

| | |
|---|---|
| How was the realism assessed externally? | __________________________ |

**Other limitations or requirements for the generated synthetic data**

| | |
|---|---|
| Does the method have other limitations or requirements for synthetic data that have not already been mentioned? | ◯ Yes<br>◯ No |
| Describe the requirements / limitations regarding to synthetic data | __________________________ |

**Advantages and disadvantages of the method**

| | |
|---|---|
| Did the authors discuss the advantages / disadvantages of the method? | ◯ Yes<br>◯ No |
| Write down the advantages of the method according to the authors | __________________________ |
| Write down the disadvantages of the method according to the authors | __________________________ |
| Write down the advantages of the method according to you | __________________________ |
| Write down the disadvantages of the method according to you | __________________________ |

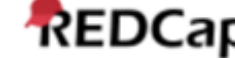

*A.3.4 Assessment of bias and reporting quality*

Page 1

# Assessment of bias and reporting quality

Please complete the survey below.

For more information, see "Risk of bias in individual studies" in the review protocol.

## Selection bias

| | |
|---|---|
| Does the study show evidence of selection bias?<br><br>Assumption: The data used and the choice of model(s) should always be justified.<br><br>Examples:<br><br>Using a data set that is known in advance to perform poorly with another method that is used as a reference for the developed method Post hoc alteration of data or model inclusion based on arbitrary or subjective reasons Using different training, validation, or test sets when evaluating the method performance | ◯ Yes<br>◯ No<br>◯ Possibly<br>(The option "Possibly" can be used in a situation where there is no clear evidence of bias, but there is something to point out about the subject.) |
| Describe the (possible) selection bias present | ______________________________ |

## Performance bias

| | |
|---|---|
| Does the study show evidence of performance bias?<br><br>Assumption: Method comparison procedures should be fair and carefully described.<br><br>Examples:<br><br>No fine-tuning is performed on the reference methods while the method in question is fine-tuned. | ◯ Yes<br>◯ No<br>◯ Possibly<br>(The option "Possibly" can be used in a situation where there is no clear evidence of bias, but there is something to point out about the subject.) |
| Describe the (possible) performance bias present | ______________________________ |
| In how many comparisons out of all reported comparisons did the method perform worse than another comparison method. | ________________________<br>(Give a fraction worse/total or write 0/1 if the method performed best/worst in every comparison) |
| List the situations in which the method performed worse than the other methods or write all and give the amount of comparisons reported, if the method performed worst in every comparison. | ______________________________ |

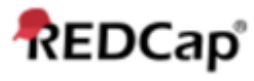

## Reporting bias

Does the study show evidence of reporting bias?

Assumption: All metrics used in the study to evaluate the performance of the method should be described in the study and the results for these should be available to the reader.

Examples:

The performance of the method has been found to be measured in some way, but the results are only partially or not at all presented.

◯ Yes
◯ No
◯ Possibly
(The option "Possibly" can be used in a situation where there is no clear evidence of bias, but there is something to point out about the subject.)

Describe the (possible) reporting bias present

__________________________________

## Inconsistency, imprecision and indirectness of reporting

Did the study show evidence of

☐ Inconsistency of reporting
☐ Imprecision of reporting
☐ Indirectness of reporting
☐ None of the above

Describe the type of inconsistency present

__________________________________

Describe the type of imprecision present

__________________________________

Describe the type of indirectness present

__________________________________

## Competing interests

Were competing interests reported?

◯ Yes
◯ No
◯ Not available

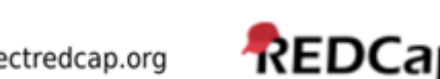

### A.4 Risk of bias and reporting quality assessment

*A.4.1 Risk of bias assessment framework*

The framework used to assess the risk of bias. This table outlines different biases that may influence the evaluation of the methods' performance. The risk of these biases were assessed from the included publications. The table presents the fundamental principles (Rationale) that guided the assessment as well as the challenges involved in recognizing each type of bias (Assessment plausibility), along with illustrative examples of each type of bias.

| Bias | Rationale | Assessment plausibility | Examples |
|---|---|---|---|
| Selection bias | Assessing the method's performance requires fairness in data representation, use of suitable metrics, and equal potential across methods to perform specific tasks. This necessitates clear justifications for input data, metrics, and reference method selection. | Detecting selection bias is difficult because any assessment approaches taken prior to the final publication may not be fully disclosed, making it difficult to assess favoritism towards the primary method. The reviewers may also be unaware of instances where a particular dataset did not work well with a particular method. | - Adjusting data or models based on arbitrary factors.<br>- Using different datasets to evaluate different methods<br>- Selectively using data or methods to favor the primary method. |
| Performance bias | To ensure a fair performance evaluation across methods, it is essential that a transparent and detailed description of the comparison and training procedures has been provided. | Detecting performance bias is challenging when the model selection and training details are incomplete or not reported. It becomes possible when the authors provide these details and mention using reference methods without task optimization. | - Not giving the reference methods a fair opportunity to perform well, e.g., through intentionally inadequate model training compared to the primary method. |
| Reporting bias | To ensure research transparency, it is important that all research evaluation metrics are comprehensively documented, and the results are shared. | Detecting the bias should be straightforward when a publication or its supplementary material lacks or incompletely presents results for the evaluation approaches mentioned in the study. | - Results are either incomplete or missing |

*A.4.2 Risk of bias in individual studies (detailed explanations)*

The following table contains detailed explanations of the identified risk of bias within each study. The risk of bias was assessed using the criteria presented in Appendix A.4.1.

| Authors | Performance bias | Explanation | Reporting bias | Explanation |
|---|---|---|---|---|
| Bhanot et al. [10] | Possibly | The method was compared to other methods, but the training processes were not described. | | |
| Li et al. [54] | Possibly | The method was compared to other methods, but the training processes were not described. | Yes | Certain outcomes were exclusively or incompletely reported across methods and/or datasets. For instance, not all outcomes of t-tests were fully given, and patient trajectories were displayed only for the primary method and using only the MIMIC-III data. |
| Yu, He & Raghunathan [102] | Possibly | The method was compared to other methods, but the training processes were not described. | Yes | Certain findings, such as those shown in Table 2, pertained only to the primary method. Furthermore, the outcomes pertaining to IVEWare were excluded from the tabulated results of Tables 3 and 4. These specific outcomes were also omitted from the supplemental materials. |
| Zhang et al. [105] | | | Yes | The authors asserted in their work (page 602, top of the second column) that statistical insignificance of FPR and TPR was observed. However, we could not |

| Authors | Performance bias | Explanation | Reporting bias | Explanation |
|---|---|---|---|---|
| | | | | find information about the specific statistical test they used in this context. |
| Zhang, Yan & Malin [106] | Possibly | The method was compared to other methods, but the training processes were not described. | Yes | The primary method “Baseline + CFR + RS” was omitted from Figure 5 illustrating the drift in time. |
| Biswal et al. [11] | Possibly | The method was compared to other methods, but the training processes were not described. | Yes | In Figure 2, the VAE-Deconv component was absent. Within Figure 3, the depiction of outcomes is partial across various methods, and the rationale for excluding specific subfigures has not been presented. The evaluation of privacy remains either unaddressed or, at minimum, the outcomes pertaining to the alternative comparative methods and $EVA_c$ were absent from the presentation. |
| Gootjes-Dreesbach et al. [31] | | | Yes | Comparative analyses between the actual patients and virtual patients were only shown for the PPMI dataset. In Figure 6, the depiction of decoded real patients was missing from the subset pertaining to SP513. |
| Sood et al. [80] | | | Yes | Comparisons between synthetic and original variables were selectively delineated for a subset of the variables under consideration. |
| Sood et al. [81] | | | Yes | Comparisons between synthetic and original variables were selectively delineated for a subset of the variables under consideration. |
| Fisher et al. [27] | | | Yes | The authors had decided to confine the outcome section to a subset of data characterized as partially synthetic. Notably, some of the evaluation techniques could have been suitably extended to encompass fully synthetic data. The rationale behind this decision remains unclear. |
| El Kababji et al. [42] | Possibly | The method was compared to others, and the authors appropriately noted that some methods had limited capabilities for hyperparameter tuning. While this is a valuable observation, it is possible that these limitations could have influenced the results, potentially introducing performance bias. | | |
| Yoon et al. [100] | Possibly | The method was compared to other methods, but the training processes were not described. | | |
| Kuo et al. [47] | | | Yes | Some alternative GAN training methods were also tested, but their results were not displayed. |
| Hashemi et al. [37] | Possibly | The method was compared to other methods, but the training processes were not described. | | |
| Sun, Lin & Yan [83] | Possibly | The method was compared to other methods, but the training processes were not described. | Yes | The utility plot (Figure 4) was presented only for the primary method MSIC. The privacy plot (Figure 5) omitted certain methods (LSTM+MLP, MTGAN), and the gram probability plots also excluded some methods (LSTM+MLP, MTGAN, Med(B/W)GAN). |
| Wang & Sun [94] | | | Yes | Certain results were excluded regarding the comparison methods, such as privacy and utility (prediction) metrics. |

| Authors | Performance bias | Explanation | Reporting bias | Explanation |
|---|---|---|---|---|
| Das, Wang & Sun [20] | | | Yes | The dimension-wise probabilities for lab tests were excluded, and the patient-wise correlation figures (Figures 4 and 5) for the other methods were not shown. Additionally, Figures 6 and 9, along with Table 3, miss some of the reference methods. |
| Raab, Nowok & Dibben [72] | | | Yes | Analyses concerning the marginal distributions and the preservation of temporal correlations of discrete variables were not presented. |
| Haleem et al. [33] | Possibly | The method was compared to other methods, but the training processes were not described. | Possibly | From the provided figures on the original data, it seems that comparison results were shown for only a subset of the variables in the dataset. However, this cannot be confirmed due to the lack of a detailed description of the original data. |
| Pang et al. [68] | Possibly | The method was compared to other methods, but the training processes were not described. | Yes | Certain measurements (privacy metrics) and figures (4, 6, 7) were not provided for the comparison methods. |
| Kühnel et al. [46] | | | Yes | Some univariate comparisons were missing; specifically, Table 2 did not include all variables, and this information was also absent from the supplemental material. |

*A.4.3 Individual study reporting quality assessment (detailed explanations)*

Detailed explanations of the identified reporting quality deviations within each study are presented in the table below. **Inconsistency of reporting** refers to utilization of 1) identical terminology or notations to signify distinct phenomena and lacking clarification (e.g., using "noise" for both original data variation and additional privacy mechanism-induced noise without clear differentiation) or 2) disparate notations to represent the same phenomenon, both 1 and 2 introduce a potential risk of misunderstanding. **Imprecision of reporting** refers to the lack of precision (e.g. p-values reported with varying accuracies) or clarity in the presentation of information, which may lead to ambiguity or difficulty in understanding the reported data. **Indirectness of reporting** involves conveying information in a manner that is not straightforward or explicit, albeit to a lesser extent than observed in reporting bias, potentially requiring the reader to infer or deduce certain details. This can introduce a level of uncertainty or make the interpretation less direct.

| Authors | Inconsistency | Explanation | Imprecision | Explanation | Indirectness | Explanation |
|---|---|---|---|---|---|---|
| Belgodere et al. [8] | | | | | Yes | The exact number of variables was not clearly specified; instead, the term "about 18" was used (p. 9, right column, second-to-last paragraph) |
| Bhanot et al. [10] | | | Yes | The number of patients was not reported precisely (“The data set has over 30 K records”, page 2). | | |

| Authors | Inconsistency | Explanation | Imprecision | Explanation | Indirectness | Explanation |
|---|---|---|---|---|---|---|
| Biswal et al. [11] | | | Yes | The number of clinicians used to evaluate the realism score was not reported. Information on the minimum and maximum number of visits per patient and the minimum number of codes per visit was not given. | Yes | Full details about the presence disclosure test were not provided. Abbreviations like ELBO were unspecified. The nature of preliminary evaluations mentioned in the appendix remained unclear. |
| Das, Wang & Sun [20] | | | | | Yes | In equations (11) and (12), it was unclear how the numerator and denominator are derived from the data. |
| Gootjes-Dreesbach et al. [31] | Yes | Utilized three distinct notations for the differential privacy budget parameter. | Yes | Subfigure 9.1 did not specify the epsilon used in that figure. | | |
| Li et al. [54] | | | Yes | Statistically significant p-values were not reported as precisely as values above 0.05. | Yes | The meaning of mean and standard deviation for a discrete-valued feature was unclear (page 13, section 4.3). Figure 5 states that the y-axis represented the probability distribution of Mechanical Ventilation and Vasopressor being applied ("On"). It was unclear how the y-axis can exceed the range of [0,1]. In reference to differential privacy, it was stated that delta ≤ 0.001 (p. 21), but it was unclear what the exact delta was in each situation, e.g. if the delta remained constant for all values of epsilon presented in Figure 7b. |
| Sood et al. [80] | | | Yes | The number and types of variables employed in the actual synthesis of data remained unclear. | | |
| Sun, Lin & Yan [83] | | | | | Yes | It is unclear how the Dimwise score was computed. Additionally, the model used to assess utility wasn't specified, leaving out important details about the task and the exact model applied. |
| Wendland et al. [95] | | | Yes | The p-value on page 3, right column, first paragraph, was reported with different precision compared to the subsequent p-values (which have two significant figures). | | |

| Authors | Inconsistency | Explanation | Imprecision | Explanation | Indirectness | Explanation |
|---|---|---|---|---|---|---|
| Yoon et al. [100] | | | | | Yes | Some abbreviations, like AP, were either not explained or not introduced when first mentioned. |
| Zhang et al. [105] | | | | | Yes | The data description in Table 1 shows the gender distribution, but the article lacks clarity on whether this variable was utilized in data synthesis or analyses. |

### A.5 Study selection: excluded publications

The primary reason for exclusion was wrong data type (n = 237), mostly cross-sectional, e.g., [1, 69, 89, 104], survival [13, 24, 43, 98] or time-series data [39, 87, 92]. Publications compromising the temporal structure in longitudinal data were categorized as having wrong data type, e.g., [5, 21, 55]. Publications lacking SDG (n = 92) were typically introductions of a specific synthetic data framework, e.g., [12, 23, 57, 60] or data simulations [28, 56, 61]. Exclusions due to partially synthetic data (n = 51) were largely related to data augmentation using techniques such as Synthetic Minority Over-Sampling Technique (SMOTE) [50] or its variants, e.g., [59, 70, 78, 84, 91].

We excluded 38 publications as we could not determine their eligibility, stemming from incomplete data, incomplete method description, or restricted access to the cited references, data, or algorithms, e.g., [30, 41, 82, 103]. Additionally, 32 studies were excluded for relying solely on standard probability distributions to simulate data, e.g., [22, 38, 44, 66]. Furthermore, 20 studies were excluded for failing to acknowledge the longitudinal nature of data, e.g., [3, 4, 19, 26, 34, 53], although the original datasets included variables with repeated measurements. Lastly, we identified four duplicates and one publication of wrong literature type (thesis).

### A.6 Reference methods

The following table presents the reference methods used to benchmark the primary methods. Those reference methods included in the main review as primary methods are marked with "(included)" in the last column. Some reference methods were not included as primary methods because their original publications did not design or implement the method for longitudinal data, or we could not confirm how the authors modified the method to address the longitudinal aspect. Therefore, we excluded these methods from further inspection but included them here, as they may still be valuable for other users.

| Study | Primary method | Reference methods |
|---|---|---|
| Biswal et al. [11] | EVA | EVAc<br>biLSTM [76]<br>VAE-LSTM [14]<br>VAE-Deconv [77] |
| Das, Wang & Sun [20] | TWIN | KNN<br>EVA [11] (included)<br>SynTEG [105] (included)<br>PromptEHR [94] (included) |
| Haleem et al. [33] | TC-MultiGAN | CTGAN [97] (included)<br>Gaussian Copula [50]<br>Dragan GAN [45]<br>Cramer GAN [9] |
| Hashemi et al. [37] | TAP-GAN | TimeGAN [99] |
| Kühnel et al. [46] | VAMBN-MT | VAMBN [31] (included) |

| Study | Primary method | Reference methods |
|---|---|---|
| | | VAMBN-FT |
| Kuo et al. [47] | HGG+VAE+Buffer | Health Gym GAN (HGG) [49] (included)<br>HGG+G_EOT+VAE+Buffer |
| Li at al. [54] | EHR-M-GAN | C-RNN-GAN [62]<br>RCGAN [25]<br>TimeGAN [99]<br>MedGAN [17]<br>seqGAN[101]<br>SynTEG [105] (included)<br>DualAEE [51]<br>PrivBayes [104] |
| Lu et al. [58] | MTGAN | MedGAN [17]<br>CTGAN [97] (included)<br>EMR-WGAN [107]<br>RDP-CGAN [88]<br>WGAN-GP [32]<br>TimeGAN [99]<br>T-CGAN [75] |
| Pang et al. [68] | CEHR-GPT | CEHR-BERT<br>GPT-Vanilla<br>GPT-OUTPAT |
| Sun, Lin & Yan [83] | MSIC | MedGAN [7]<br>Med(B/W)GAN [5]<br>EVA [11] (included)<br>MTGAN [58] (included)<br>PromptEHR [94] (included) |
| Theodorou et al. [86] | HALO | HALO-coarse<br>EVA [11] (included)<br>SynTEG [105] (included)<br>LSTM [50]<br>GPT-2 [73] |
| Theodorou et al. [85] | ConSequence | HALO [86] (included)<br>Semantic Loss [96]<br>CCN [29]<br>MultiPlexNet [40]<br>SPL [2] |
| Wang & Sun [94] | PromptEHR | LSTM + MLP<br>LSTM + MedGAN [17]<br>SynTEG [105] (included)<br>GPT-2 [73] |
| Wendland et al. [95] | MultiNODEs | VAMBN [31] (included) |
| Yoon et al. [100] | EHR-Safe | TimeGAN [99]<br>RCGAN[25]<br>C-RNN-GAN [62] |
| Yu, He & Raghunathan [102] | SPMI | IVEware Version 0.3 [74]<br>Synthpop [65] (included) |

### A.7 Datasets used in the included publications

The following table lists all datasets used in the included publications to generate synthetic data.

| Datasets | Data type | Availability | Study | Subjects | Numerical variables | Categorical variables | Time-varying variables | Repeated measurements |
|---|---|---|---|---|---|---|---|---|
| MIMIC-III | Clinical database | Public | [7] | 8 260 | 9 | 1 | 9 | 5 |
| | | | [54] | 28 344 | 78 | 20 | 98 | 24 |
| | | | [58] | 7 493 | 0 | 4 880 | 4 880 | avg. 2.6 |
| | | | [48,49] | 3 910 | 9 | 13 | 20 | 48 |
| | | | [49] | 2 164 | 35 | 11 | 42 | 2–20 |
| | | | [86] | 929 268 | 0 | 9882 | 9882 | avg. 3.9 |
| | | | [86] | 46 520 | 15 | 31 | 17 | avg. 11.9 |
| | | | [94] | 46 520 | 0 | 4 | 4 | NA |
| | | | [83] | 5 794 | 0 | 3 | 3 | avg. 2.1 |
| | | | [8] | 14 681 | NA | NA | ~18 | 48 |
| | | | [85] | 46 520 | 0 | 1616 | 1610 | avg. 1.3 |
| | | | [100] | 19 946 | 79 | 11 | 84 | 1–30 |
| MIMIC-IV | Clinical database | Public | [58] | 10 000 | 0 | 6 102 | 6 102 | avg. 3.6 |
| | | | [37] | 9 133 | 6 | 0 | 6 | 5 |
| | | | [83] | 36 615 | 0 | 3 | 3 | avg. 3.5 |
| | | | [64] | 6 535 | NA | 3 | 3 | NA |
| PPMI | Patient data | Public | [80] | 362 | NA | NA | 38 | 2–12 |
| | | | [95] | 354 | 53 | 15 | 25 | 5–12 |
| | | | [31] | 557 | NA | NA | 38* | 5 |
| eICU | Clinical database | Public | [54] | 99 015 | 55 | 19 | 74 | 24 |
| | | | [100] | 198 707 | 53 | 1 | 51 | 2–50 |
| VUMC | EHR data | No | [106] | 59 617 | 0 | 1 276 | 1 276 | 25–200 |
| | Synthetic derivate | No | [105] | 2 187 629 | 0 | 1 799 | 1 799 | avg.12.1 |
| ADNI | Patient data | Public | [80] | 689 | NA | NA | 18 | 4 |
| Alberta Health | EHR data | Upon request | [63] | 100 000 | NA | NA | 6 | max. 1000 |
| All of Us | EHR data | Public | [106] | 59 617 | 0 | 526 | 526 | 10–200 |
| ASD | Health data | No | [10] | > 280 000 | 7 | 2 | 7 | 10 |
| ATUS | Behavioral data | Public | [10] | > 30 000 | 1 | 4 | 1 | 30 |
| Breast cancer | Clinical trial data | Upon request | [20] | 971 | 0 | 3 | 3 | max. 14 |
| CCTG MA27 | Clinical trial data | No | [42] | 7 576 | NA | NA | NA | NA |
| CDC | EHR data | No | [93] | 9 298 | 1 | 100 | 88 | 2–11 |
| CODR-AD | Clinical database | No | [27] | 1 909 | 38 | 6 | 36 | 7 |
| CUIMC New York Presbyterian | EHR data | No | [68] | ~ 2 300 000 | 3 | 8 | 5 | 2–102 |
| DONALD | Health data | No | [46] | 1312 | 28 | 7 | 33 | 16 |
| GATEKEEPER | Health data | No | [33] | 86 | NA | NA | NA | max. 24 |
| HiRID | Clinical database | Public | [54] | 14 129 | 50 | 39 | 89 | 24 |
| HIV | EuResist integrated database | Public | [47–49] | 8 916 | 3 | 12 | 13 | 10–100 |
| HRS | Longitudinal survey | No | [102] | 12 652 | 7 | 41 | 11 | 2–3 |

| Datasets | Data type | Availability | Study | Subjects | Numerical variables | Categorical variables | Time-varying variables | Repeated measurements |
|---|---|---|---|---|---|---|---|---|
| Multi-census | Census data | No | [90] | NA | NA | NA | NA | NA |
| NACC | Patient data | Public | [95] | 2 284 | 4 | 3 | 3 | 4 |
| NSABP B34 | Clinical trial data | No | [42] | 3 323 | NA | NA | NA | NA |
| PAMF EHR | EHR data | No | [11] | 258 555 | 0 | 10 437 | 10 437 | avg. 53.8 |
| REaCT | Clinical trial data | No | [42] | 48–230 | NA | NA | NA | NA |
| SIPP | Longitudinal survey | Yes | [15] | 23 374 | 0 | 1 | 1 | 12 |
| Small cell lung carcinoma | Clinical trial data | Upon request | [20] | 77 | 0 | 3 | 3 | max. 5 |
| SP513 | Clinical trial data | No | [31] | 560 | NA | NA | 35* | 2–11* |
| SPRINT | Clinical trial data | No | [7] | 6 502 | 3 | 1 | 3 | 12 |
| Status File | Employment data | No | [6] | 3 511 824 | 5 | 24 | 22 | 24 |
| SWOG 0307 | Clinical trial data | No | [42] | 6 097 | NA | NA | NA | NA |
| TREND | Patient data | No | [81] | 1 178 | 1 | 18 | 18 | 4 |
| UK LS | Census data | No | [72] | > 186 000 | 1 | 4 | 5 | 2 |
| Unnamed | Patient data | No | [79] | 580 000 | NA | NA | 2 | 2 |
| US claims data | Insurance data | No | [85] | 1 006 321 | 0 | 1824 | 1817 | avg. 35.4 |

NA: not available; EHR: electronic health records; avg.: average; max.: maximum; ~: approximately; *: calculated from presented materials by the corresponding author of this systematic literature review

## A.8 PRISMA checklist

PRISMA 2020 Checklist

| Section and Topic | Item # | Checklist item | Location where item is reported |
|---|---|---|---|
| **TITLE** | | | |
| Title | 1 | Identify the report as a systematic review. | Title |
| **ABSTRACT** | | | |
| Abstract | 2 | See the PRISMA 2020 for Abstracts checklist. | Abstract |
| **INTRODUCTION** | | | |
| Rationale | 3 | Describe the rationale for the review in the context of existing knowledge. | Section 1.1 |
| Objectives | 4 | Provide an explicit statement of the objective(s) or question(s) the review addresses. | Section 1.2 |
| **METHODS** | | | |
| Eligibility criteria | 5 | Specify the inclusion and exclusion criteria for the review and how studies were grouped for the syntheses. | Section 2.1 |
| Information sources | 6 | Specify all databases, registers, websites, organisations, reference lists and other sources searched or consulted to identify studies. Specify the date when each source was last searched or consulted. | Section 2.2 |
| Search strategy | 7 | Present the full search strategies for all databases, registers and websites, including any filters and limits used. | Section 2.3 + A.1 |
| Selection process | 8 | Specify the methods used to decide whether a study met the inclusion criteria of the review, including how many reviewers screened each record and each report retrieved, whether they worked independently, and if applicable, details of automation tools used in the process. | Section 2.4 + A.2 |
| Data collection process | 9 | Specify the methods used to collect data from reports, including how many reviewers collected data from each report, whether they worked independently, any processes for obtaining or confirming data from study investigators, and if applicable, details of automation tools used in the process. | Section 2.5 |
| Data items | 10a | List and define all outcomes for which data were sought. Specify whether all results that were compatible with each outcome domain in each study were sought (e.g. for all measures, time points, analyses), and if not, the methods used to decide which results to collect. | Section 2.5 + A.3 |
| | 10b | List and define all other variables for which data were sought (e.g. participant and intervention characteristics, funding sources). Describe any assumptions made about any missing or unclear information. | Section 2.5 + A.3 |
| Study risk of bias assessment | 11 | Specify the methods used to assess risk of bias in the included studies, including details of the tool(s) used, how many reviewers assessed each study and whether they worked independently, and if applicable, details of automation tools used in the process. | Section 2.6. + A.4 |
| Effect measures | 12 | Specify for each outcome the effect measure(s) (e.g. risk ratio, mean difference) used in the synthesis or presentation of results. | Not applicable |
| Synthesis methods | 13a | Describe the processes used to decide which studies were eligible for each synthesis (e.g. tabulating the study intervention characteristics and comparing against the planned groups for each synthesis (item #5)). | Section 2.7 |
| | 13b | Describe any methods required to prepare the data for presentation or synthesis, such as handling of missing summary statistics, or data conversions. | Section 2.5<br>Section 2.7 |
| | 13c | Describe any methods used to tabulate or visually display results of individual studies and syntheses. | Section 2.7 |
| | 13d | Describe any methods used to synthesize results and provide a rationale for the choice(s). If meta-analysis was performed, describe the model(s), method(s) to identify the presence and extent of statistical heterogeneity, and software package(s) used. | Section 2.7 |
| | 13e | Describe any methods used to explore possible causes of heterogeneity among study results (e.g. subgroup analysis, meta-regression). | Not applicable |
| | 13f | Describe any sensitivity analyses conducted to assess robustness of the synthesized results. | Not applicable |
| Reporting bias assessment | 14 | Describe any methods used to assess risk of bias due to missing results in a synthesis (arising from reporting biases). | Not applicable |
| Certainty assessment | 15 | Describe any methods used to assess certainty (or confidence) in the body of evidence for an outcome. | Not applicable |

| Section and Topic | Item # | Checklist item | Location where item is reported |
|---|---|---|---|
| **RESULTS** | | | |
| Study selection | 16a | Describe the results of the search and selection process, from the number of records identified in the search to the number of studies included in the review, ideally using a flow diagram. | Section 3.1 |
| | 16b | Cite studies that might appear to meet the inclusion criteria, but which were excluded, and explain why they were excluded. | A.4 |
| Study characteristics | 17 | Cite each included study and present its characteristics. | Section 3.2 |
| Risk of bias in studies | 18 | Present assessments of risk of bias for each included study. | Section 3.3 + A.4 |
| Results of individual studies | 19 | For all outcomes, present, for each study: (a) summary statistics for each group (where appropriate) and (b) an effect estimate and its precision (e.g. confidence/credible interval), ideally using structured tables or plots. | Not applicable |
| Results of syntheses | 20a | For each synthesis, briefly summarise the characteristics and risk of bias among contributing studies. | Section 3 + 4 |
| | 20b | Present results of all statistical syntheses conducted. If meta-analysis was done, present for each the summary estimate and its precision (e.g. confidence/credible interval) and measures of statistical heterogeneity. If comparing groups, describe the direction of the effect. | Section 3.4<br>Section 3.5 |
| | 20c | Present results of all investigations of possible causes of heterogeneity among study results. | Section 4 |
| | 20d | Present results of all sensitivity analyses conducted to assess the robustness of the synthesized results. | Not applicable |
| Reporting biases | 21 | Present assessments of risk of bias due to missing results (arising from reporting biases) for each synthesis assessed. | Not applicable |
| Certainty of evidence | 22 | Present assessments of certainty (or confidence) in the body of evidence for each outcome assessed. | Section 4 |
| **DISCUSSION** | | | |
| Discussion | 23a | Provide a general interpretation of the results in the context of other evidence. | Section 4 |
| | 23b | Discuss any limitations of the evidence included in the review. | Section 4.1 |
| | 23c | Discuss any limitations of the review processes used. | Section 4.1 |
| | 23d | Discuss implications of the results for practice, policy, and future research. | Section 4.2 |
| **OTHER INFORMATION** | | | |
| Registration and protocol | 24a | Provide registration information for the review, including register name and registration number, or state that the review was not registered. | Acknowledgements |
| | 24b | Indicate where the review protocol can be accessed, or state that a protocol was not prepared. | Acknowledgements |
| | 24c | Describe and explain any amendments to information provided at registration or in the protocol. | Acknowledgements |
| Support | 25 | Describe sources of financial or non-financial support for the review, and the role of the funders or sponsors in the review. | Acknowledgements |
| Competing interests | 26 | Declare any competing interests of review authors. | Title page; no COI |
| Availability of data, code and other materials | 27 | Report which of the following are publicly available and where they can be found: template data collection forms; data extracted from included studies; data used for all analyses; analytic code; any other materials used in the review. | Acknowledgements |

*From:* Page MJ, McKenzie JE, Bossuyt PM, Boutron I, Hoffmann TC, Mulrow CD, et al. The PRISMA 2020 statement: an updated guideline for reporting systematic reviews. BMJ 2021;372:n71. doi: 10.1136/bmj.n71

For more information, visit: http://www.prisma-statement.org/

## APPENDIX REFERENCES